\documentclass[twocolumn]{aastex61}
\hypersetup{
  pdfinfo={Title={Matsuno2018a.pdf},Author={Tadafumi Matsuno}}}
\begin{document}
\title{Optical high-resolution spectroscopy of 14 young $\alpha$-rich stars \footnote{The data presented herein were obtained at the W. M. Keck Observatory, which is operated as a scientific partnership among the California Institute of Technology, the University of California and the National Aeronautics and Space Administration. The Observatory was made possible by the generous financial support of the W. M. Keck Foundation.}}
\author{Tadafumi Matsuno}
\affiliation{Department of Astronomical Science, School of Physical Sciences, SOKENDAI (The Graduate University for Advanced Studies), Mitaka, Tokyo 181-8588, Japan}
\email{tadafumi.matsuno@nao.ac.jp}
\affiliation{National Astronomical Observatory of Japan (NAOJ), 2-21-1 Osawa, Mitaka, Tokyo 181-8588, Japan}
\author{David Yong}
\affiliation{Research School of Astronomy and Astrophysics, Australian National University, Canberra, ACT 2611, Australia}
\author{Wako Aoki}
\affiliation{National Astronomical Observatory of Japan (NAOJ), 2-21-1 Osawa, Mitaka, Tokyo 181-8588, Japan}
\affiliation{Department of Astronomical Science, School of Physical Sciences, SOKENDAI (The Graduate University for Advanced Studies), Mitaka, Tokyo 181-8588, Japan}
\author{Miho N. Ishigaki}
\affiliation{Kavli Institute for the Physics and Mathematics of the Universe (WPI), The University of Tokyo, Kashiwa, Chiba 277-8583, Japan}
\begin{abstract}
We report chemical abundances of 14 young $\alpha$-rich stars including neutron-capture elements based on high-quality optical spectra from HIRES/Keck~I and differential line-by-line analysis.
From the comparison of the abundance patterns of young $\alpha$-rich stars to those of nearby bright red giants with a similar metallicity range ($-0.7<[\mathrm{Fe/H}]<-0.2$), we confirm their high $\alpha$-element abundances reported by previous studies based on near-infrared spectroscopy. 
We reveal for the first time low abundances of $s$-process elements and high abundances of $r$-process elements. 
All the abundances are consistent with those seen in typical $\alpha$-rich population of the Galactic disk, and no abundance anomalies are found except for Li-enhancement in one object previously reported and mild enhancement of Na in two stars.
In particular, the lack of $s$-process enhancement excludes the hypothesis that mass transfer from asymptotic giant branch stars plays an important role in the formation of young $\alpha$-rich stars.
The high frequency of radial velocity variation (more than $50\%$) is also confirmed.
We argue that mass transfer from low-mass red giants is the likely dominant formation mechanism for young $\alpha$-rich stars.
\end{abstract}
\section{Introduction}
Despite the importance of stellar ages in astrophysics, their determination has been difficult in many cases \citep[for review, see ][]{Soderblom2010}.
Recent space telescopes have revolutionized this situation by providing precise long term light curves of red giants.
Fourier transform of the light curves enables us to measure the oscillation of stars, which reflect fundamental properties of stars including stellar mass.
This approach is called asteroseismology \citep{Aerts2010}. 
Thanks to CoRoT and \textit{Kepler} observations, masses of a large number of red giants are estimated for the first time, which can then be converted to stellar ages through stellar evolution models \citep[e.g., ][]{Chaplin2011,Miglio2013}.

\citet{Martig2015a} and \citet{Chiappini2015} combined these asteroseismic ages with a large spectroscopic survey, Apache Point Observatory Galactic Evolution Experiment \citep[APOGEE; ][]{Majewski2017} within the Sloan Digital Sky Survey \citep[SDSS; ][]{York2000}, and found a small and unexpected population in the Galactic disk, so-called young $\alpha$-rich stars that have high [{$\alpha$}/{Fe}] and masses of about $1.5\,\mathrm{M_{\odot}}$.
A majority of stars in the Galactic disk show a clear age-[{$\alpha$}/{Fe}] trend, such that older stars have higher [{$\alpha$}/{Fe}].
This tendency is well explained by simple galactic chemical evolution models, in which delayed enrichment of Fe by type Ia supernovae lowers the [{$\alpha$}/{Fe}] ratio as time passes \citep[e.g., ][]{Tinsley1979,Matteucci1986}.
The property of young $\alpha$-rich stars is contrary to this tendency: they have high [{$\alpha$}/{Fe}], but are estimated to be young from asteroseismology.
The fraction of such stars is estimated to be $\sim6\,\%$ among the $\alpha$-rich population \citep{Martig2015a}.

One possible explanation for young $\alpha$-rich stars is that their masses do not reflect ages but are affected by mass acquisition events after star formation \citep[evolved blue straggler scenario, ][]{Martig2015a,Chiappini2015,Yong2016,Jofre2016,Izzard2018}.
The basic idea is that these young $\alpha$-rich stars are evolved counterparts of blue stragglers, which can usually be identified as a minor component of stars in stellar clusters.
Blue stragglers are bluer and brighter than turn-off stars in stellar clusters, and hence, to be more massive than the rest of stars in the cluster.
They are considered as a result of stellar merger or mass transfer in binary systems \citep{McCrea1964,Hills1976}.

Another explanation is that young $\alpha$-rich stars are formed in a special place in the Galaxy, such as the end point of the Galactic bar (peculiar formation site scenario).
\citet{Chiappini2015} found that young $\alpha$-rich stars are preferentially found in the inner region of the Galaxy and with distinct kinematics from other $\alpha$-rich stars.
Although \citet{Martig2015a} could not confirm these findings, they showed that the fraction of young $\alpha$-rich stars among $\alpha$-rich stars is higher than the fraction of blue stragglers in stellar clusters, which might support peculiar formation site scenario.
 
Although APOGEE has provided high-resolution ($R\sim 22500$) H-band spectra, from which chemical abundances can be derived for $\alpha$-elements and iron-group elements, detailed information of chemical abundances is still limited. 
One of the limitations is that we are unable to measure abundances of many neutron-capture elements from infrared spectra. 
Currently only four young $\alpha$-rich stars have measurements of neutron-capture elements from optical spectra \citep{Yong2016,Jofre2015}. 

An advantage of measuring neutron-capture elements is that they are produced in different sites with different timescales from $\alpha$-elements or iron-group elements.
Therefore, they provide independent information on the chemical enrichment history of the population of young $\alpha$-rich stars.
Moreover, the signature of mass transfer from a companion in a binary system can be seen in the abundance pattern of neutron-capture elements in some cases (e.g., Ba stars). 
Although the previously investigated four stars do not show clear anomalies, a correlation between Ba abundance and mass among young $\alpha$-rich stars  is suggested by \citet{Yong2016}. 
Since the sample size is still small, further studies of neutron-capture elements with high-precision analysis ($\sigma([\mathrm{X/Fe}]\lesssim 0.1$) for a larger sample are desired to investigate the possibility of $s$-process enhancement and the trend with mass.

Another constraint on the origin of young $\alpha$-rich stars is obtained from variation of radial velocity, which is a signature of the existence of a companion and supports the evolved blue straggler scenario.
While this approach was conducted by \citet{Jofre2016}, additional measurements provide stronger constraints.

The aim of this paper is to obtain precise abundances including neutron-capture elements from optical high-resolution spectra. 
In Section \ref{obs}, we describe observations and data reduction.
The procedure of the abundance analysis is explained in Section \ref{ab}. 
Based on the results presented in Section \ref{results}, we finally discuss the origin of young $\alpha$-rich stars in Section \ref{discussion}.

\section{Observation and data reduction\label{obs}}
\begin{deluxetable*}{lrrrrrr}
  \tablecaption{Observation log \label{tableobs}}
  \tablehead{ 
   \colhead{Object}                                & \colhead{2MASS ID} & \colhead{Exp. Time.}    & \colhead{SNR at $6000\,\mathrm{\AA}$} & \colhead{Mass}    & \colhead{$RV$}   & \colhead{$RV$ (APOGEE)}             \\
                                                   &                     & \dcolhead{(\mathrm{s})} & \dcolhead{(\mathrm{pix^{-1}})\tablenotemark{a}}        & \dcolhead{(\mathrm{M_{\odot}})} & \dcolhead{(\mathrm{km\,s^{-1}})} & \dcolhead{(\mathrm{km\,s^{-1}})}  
   }
   \startdata
   \multicolumn{7}{c}{Young $\alpha$-rich stars}\\
CoRoT 101364068      & J19281113--0020004    & 9000 &  268 & $1.30 \pm 0.15$ &  -49.3   & -45.6    \\
CoRoT 101665008      & J19302198+0018463     & 9000 &  252 & $1.28 \pm 0.15$ &  0.1     & -0.0     \\
CoRoT 101748322      & J19305707--0008228    & 3600 &  206 & $1.34 \pm 0.11$ &  20.1    & 11.7     \\
KIC 10525475         & J19102133+4743193     &  900 &  214 & $1.43 \pm 0.18$ &  -44.6   & -39.6    \\
KIC 11394905         & J19093999+4913392     & 1800 &  249 & $1.40 \pm 0.18$ &  -74.7   & -75.4    \\
KIC 11445818         & J19052620+4921373     & 3600 &  363 & $1.49 \pm 0.16$ &  -26.9   & -26.5    \\
KIC 11823838         & J19455292+5002304     & 1200 &  180 & $1.57 \pm 0.18$ &  -27.0   & -18.1    \\
KIC 3455760          & J19374569+3835356     & 1200 &  145 & $1.49 \pm 0.16$ &  -46.5   & -47.6    \\
KIC 3833399          & J19024305+3854594     &  300 &  266 & $1.45 \pm 0.17$ &  -61.0   & -62.0    \\
KIC 4143460          & J19101154+3914584     & 2310 &  229 & $1.58 \pm 0.20$ &  6.3     & 6.5      \\
KIC 4350501          & J19081716+3924583     & 1800 &  191 & $1.65 \pm 0.20$ &  -83.1   & -83.3    \\
KIC 5512910          & J18553092+4042447     & 8100 &  169 & $1.66 \pm 0.22$ &  -40.4   & -39.1    \\
KIC 9269081          & J19032243+4547495     & 2700 &  282 & $2.06 \pm 0.43$ &  -86.6   & -85.8    \\
KIC 9821622          & J19083615+4641212     & 2700 &  261 & $1.71 \pm 0.26$ &  -5.3    & -5.3     \\
   \multicolumn{7}{c}{Comparison stars}\\
2M0001+2415          & J00014289+2415111     &  40  &  304 & $1.40 \pm 0.44$ & -7.1    & -7.2    \\
2M0006+4053          & J00062019+4053555     &  40  &  230 & $0.79 \pm 0.23$ & -76.0   & -75.4   \\
2M0040+5927          & J00402003+5927517     &  60  &  214 & $0.76 \pm 0.24$ & -12.5   & -12.4   \\
2M0040--0421         & J00404236--0421065    &  20  &  228 & $1.49 \pm 0.44$ & 38.0    & 37.3    \\
2M0049+4104          & J00491615+4104545     &  40  &  307 & $1.67 \pm 0.54$ & -46.6   & -46.7   \\
2M0158+7622          & J01580554+7622122     &  80  &  210 & $0.70 \pm 0.22$ & -46.1   & -45.8   \\
2M0240+0253          & J02404734+0253546     &  60  &  244 & $0.85 \pm 0.27$ & 69.1    & 71.8    \\
2M0248+1817          & J02483208+1817018     &  30  &  325 & $1.60 \pm 0.48$ & 47.1    & 46.7    \\
2M0328+3548          & J03284901+3548266     &  60  &  247 & $0.68 \pm 0.21$ & 8.7     & 8.1     \\
2M0419+1416          & J04194460+1416257     &  60  &  222 & $0.84 \pm 0.25$ & -18.6   & -19.1   \\
2M2114+3914          & J21142354+3914355     &  90  &  245 & $0.75 \pm 0.24$ & -41.3   & -41.4   \\
2M2119+5303          & J21194076+5303290     & 150  &  440 & $1.90 \pm 0.59$ & 11.1    & 11.2    \\
2M2156+2109          & J21563597+2109405     &  40  &  369 & $1.57 \pm 0.50$ & -24.6   & -24.4   \\
2M2228+2701          & J22281112+2701075     &  30  &  371 & $2.65 \pm 0.77$ & -39.9   & -39.6   \\
2M2308+0207          & J23084093+0207404     &  20  &  346 & $1.40 \pm 0.41$ & -13.6   & -25.2   \\
2M2344+5547          & J23444837+5547589     &  20  &  254 & $1.64 \pm 0.49$ & 5.5     & 6.1     \\
  \enddata 
\tablenotetext{a}{Each pixel corresponds to $1.33\,\mathrm{km\,s^{-1}}$.}
\end{deluxetable*}
Fourteen young $\alpha$-rich stars are selected from \citet{Martig2015a} and \citet{Chiappini2015}.
All the selected stars have effective temperature ($T_{\rm eff}$) between $4500\,\mathrm{K}$ and $5000\,\mathrm{K}$ according to their estimates.
The narrow range of $T_{\rm eff}$ enables us to achieve high precision in deriving relative abundances between stars.
In addition, 16 nearby bright giants in Hipparcos/APOGEE sample of \citet{Feuillet2016} are selected as typical disk stars to compare the abundance pattern of young $\alpha$-rich stars.
The comparison sample covers similar range in $T_{\rm eff}$ and [{Fe}/{H}] as the main targets ($4500\,\mathrm{K}\lesssim T_{\rm eff} \lesssim 5000\,\mathrm{K}$, $-0.7\lesssim [\mathrm{Fe/H}]\lesssim -0.2$). 
\added{We try to include both $\alpha$-poor and $\alpha$-rich disk stars in the comparison sample, among which five result in being $\alpha$-rich, though they are inclined to be metal-rich (see results).} 
Although the selection does not include age, it is highly unlikely that the comparison sample includes many young $\alpha$-rich stars given their small fraction \replaced{in surveys}{among $\alpha$-rich stars (see Introduction)}.

Observations were conducted on August 7 and 8, 2016, under good sky condition with HIRES \citep{Vogt1994} on Keck I telescope through the time exchanging program between Keck and the Subaru Telescope (proposal ID: S16B-084).
We adopt the B2 setting ($R\sim 67,000$), with the echelle angle of $-0.13^\circ$, and the red cross disperser at the angle of $0.504^\circ$.
While the obtained spectra cover from $4200\,\mathrm{\AA}$ to $8750\,\mathrm{\AA}$, we refrain from analysing bluer part of the spectra than $5200\,\mathrm{\AA}$ to avoid the effect of significant line blending. 
The read-out of the CCD data were conducted with the default $2\times 1$ binning and in the low gain setting.

Data reduction is conducted in a standard manner using \texttt{MAKEE}\footnote{\url{http://www.astro.caltech.edu/\%7etb/makee/}} version 5.2.4 including baseline subtraction, flat fielding, aperture extraction, cosmic ray rejection, and wavelength calibration.
We then shift the spectra to the rest frame with \texttt{dopcor} after measuring radial velocities using \texttt{fxcor} in IRAF \footnote{IRAF is distributed by the National Optical Astronomy Observatories, which are operated by the Association of Universities for Research in Astronomy, Inc., under cooperative agreement with the National Science Foundation.}.
The Arcturus spectrum of \citet{Hinkle2000} is used as the reference.
While the above procedure typically gives $\sim 0.05 \,\mathrm{km\,s^{-1}}$ as the uncertainties, we need to take into account the effects of temperature variation in the instrument during a night. 
The temperature variation can affect radial velocity measurements at the level of $\sim 0.5\,\mathrm{km\,s^{-1}}$ \citep{Griest2010}.  

Continuum placements are carried out by creating a continuum mask. 
We first run the \texttt{continuum} task for the spectrum of 2M0248+1817 since it has high photon counts and shows strongest absorption features.
Continuum of the other objects are placed by fitting the wavelength regions that are used in the continuum fitting for 2M0248+1817.

Information of targets and observation is summarised in Table \ref{tableobs}.

\section{Abundance analysis\label{ab}}
The line data used in this study are listed in Table \ref{tablelinelist}.
We also include effects of hyperfine structure splitting in the analysis for Sc~I, Sc~II, V~I, Mn~I, Co~I, Cu~I, Ba~II, La~II, and Eu~II.
Isotopic shifts are also included for Ba and Eu, assuming the solar ratio \citep{Asplund2009,Rosman1998}.
Line positions and relative strengths are taken from \citet{McWilliam1998} for Ba~II, \citet{Ivans2006} for La~II and Eu~II, and Robert L. Kurucz's linelist\footnote{\url{http://kurucz.harvard.edu/linelists.html}} for the others.

\begin{deluxetable*}{lrrrrrrr}
  \tablecaption{Line list, equivalent widths and FWHM \label{tablelinelist}}
\tablehead{
\colhead{Object}  & \colhead{species} & \colhead{wavelength} & \dcolhead{\chi}   & \dcolhead{\log gf}       & \dcolhead{EW}     & \dcolhead{FWHM}       & \colhead{Reference \tablenotemark{a}}    \\
                   &                   & \dcolhead{(\mathrm{\AA})} & \dcolhead{(\mathrm{eV})}  &                        & \dcolhead{(\mathrm{m\AA})} & \dcolhead{(\mathrm{\AA})} &                     
}
\startdata
2M0001+2415     &NaI   &   6154.225 &  2.102 & -1.547 &   87.5 &  0.162 &   1 \\
2M0001+2415     &NaI   &   6160.747 &  2.102 & -1.246 &   95.8 &  0.165 &   1 \\
2M0001+2415     &MgI   &   7387.689 &  5.753 & -1.000 &   84.4 &  0.210 &   1 \\
2M0001+2415     &MgI   &   7691.553 &  5.753 & -0.783 &   99.8 &  0.223 &   1 \\
2M0001+2415     &MgI   &   8717.815 &  5.933 & -0.930 &   82.9 &  0.274 &   2 \\
2M0001+2415     &AlI   &   5557.059 &  3.143 & -2.371 &   30.5 &  0.182 &   3 \\
\enddata
\tablenotetext{}{Portion of the table is shown.}
\tablenotetext{a}{
1: \citet{Kelleher2008}, 2: \citet{Rhodin2017}, 3: \citet{Kelleher2008a}, 4: \citet{Kelleher2008b}, 5: \citet{Nandy2012}, 6: \citet{Smith1981}, 7: \citet{Smith1988}, 8: \citet{Aldenius2009}, 9: \citet{Lawler1989}, 10: \citet{Lawler2013}, 11: \citet{Wood2013}, 12: \citet{Lawler2014}, 13: \citet{Sobeck2007}, 14: \citet{Lawler2017}, 15: \citet{Booth1984}, 16: \citet{DenHartog2011}, 17: \citet{Ruffoni2014}, 18: \citet{Bard1991}, 19: \citet{Bard1994}, 20: \citet{OBrian1991}, 21: \citet{Blackwell1979}, 22: \citet{Blackwell1980}, 23: \citet{Blackwell1982}, 24: \citet{Blackwell1982a}, 25: \citet{Blackwell1986}, 26: \citet{Melendez2009}, 27: \citet{Lawler2015}, 28: \citet{Wood2014}, 29: \citet{Kock1968}, 30: \citet{Hannaford1982}, 31: \citet{Klose2002}, 32: \citet{Lawler2001}, 33: \citet{Lawler2009}, 34: \citet{DenHartog2003}, 35: \citet{Lawler2001a}.
}
\end{deluxetable*}

Equivalent widths (EWs) are measured in a consistent manner for all the stars.
We include Voigt profile for the fitting of strong lines that have $\log(\mathrm{EW}/\lambda)>-5.0$, where $\lambda$ is the wavelength of the line center.

The subsequent analysis is based on line-by-line differential analysis using q$^2$ \citep{Ramirez2014}, which carries out abundance calculation using \texttt{MOOG} \citep{Sneden1973}.
Abundances are calculated with the ATLAS9 ODFNEW model atmospheres under 1D/LTE approximation with $\alpha$-enhancements \citep{Castelli2004}. 
Since the temperature range of our targets is narrow, we assume departures from the 1D/LTE approximation do not significantly affect the relative scale.
We have also carried out the same analysis with ATLAS9 ODFNEW model atmospheres without $\alpha$-enhancements and obtained consistent results.

The reference star adopted in the analysis is KIC~11445818, which has typical stellar parameters among the observed stars. 
The stellar parameters and abundances of this star set the absolute scale of our analysis and need to be fixed in advance.
The $T_{\rm eff}$ is taken from the calibrated APOGEE DR12 catalog.
The $\log g$ is calculated using a scaling relation of asteroseismology, $g/g_\odot= \nu_{\rm max}/\nu_{\mathrm{max},\,\odot} (T_{\rm eff}/T_{\mathrm{eff},\,\odot})^{0.5}$, where $\log g_\odot = 4.438,\,\nu_{\mathrm{max},\,\odot}= 3090\,\mathrm{\mu Hz},\,\mathrm{and}\,T_{\mathrm{eff},\,\odot}= 5777\,\mathrm{K}$.
The microturbulent velocity ($v_t$) is set to minimize the correlation between reduced equivalent widths ($\log (EW/\lambda)$) and abundances derived for individual neutral iron lines.

Stellar parameters of the other stars relative to KIC~11445818 are determined through fully spectroscopic analysis in this paper.
We first derive abundance for each iron line.
The stellar parameter determination process is to minimize the correlation of derived abundances, relative to the reference star, to line strengths and to excitation potentials for neutral iron lines, and the abundance difference between neutral iron lines and singly ionized iron lines.
More details can be found in Appendix.
Results with uncertainties are given in Table \ref{tableparam}.

\begin{deluxetable*}{lrrrrrr}
  \tablecaption{Stellar parameters \label{tableparam}}
  \tablehead{ 
   \colhead{Object} & \dcolhead{T_{\rm eff}} & \dcolhead{\sigma(T_{\rm eff})}  & \dcolhead{\log g} & \dcolhead{\sigma(\log g)} & \dcolhead{v_t} & \dcolhead{\sigma(v_t)} \\ 
  & \colhead{(K)} & \colhead{(K)} & \colhead{(dex)} & \colhead{(dex)} & \dcolhead{(\mathrm{km\,s^{-1}})} & \dcolhead{(\mathrm{km\,s^{-1}})}
   }
  \startdata
2M0001+2415          &4474 &   27&  1.95 &  0.09&  1.83 &  0.04 \\
2M0006+4053          &4712 &   39&  2.67 &  0.09&  1.74 &  0.04 \\
2M0040+5927          &4674 &   37&  2.36 &  0.13&  1.96 &  0.05 \\
2M0040-0421          &4561 &   41&  1.74 &  0.13&  1.93 &  0.04 \\
2M0049+4104          &4812 &   41&  2.66 &  0.10&  1.72 &  0.05 \\
2M0158+7622          &4716 &   42&  2.60 &  0.10&  1.55 &  0.05 \\
2M0240+0253          &4733 &   39&  2.50 &  0.11&  1.63 &  0.04 \\
2M0248+1817          &4440 &   32&  1.85 &  0.12&  1.88 &  0.06 \\
2M0328+3548          &4639 &   55&  2.66 &  0.14&  1.88 &  0.07 \\
2M0419+1416          &4902 &   52&  3.30 &  0.14&  1.62 &  0.07 \\
2M2114+3914          &4513 &   46&  2.34 &  0.15&  1.89 &  0.07 \\
2M2119+5303          &4764 &   35&  2.46 &  0.10&  1.76 &  0.04 \\
2M2156+2109          &4812 &   34&  2.49 &  0.09&  1.92 &  0.04 \\
2M2228+2701          &4807 &   31&  2.27 &  0.11&  1.79 &  0.04 \\
2M2308+0207          &4863 &   51&  2.36 &  0.17&  1.75 &  0.05 \\
2M2344+5547          &4883 &   30&  2.41 &  0.11&  1.81 &  0.04 \\
CRT101364068         &4618 &   48&  2.08 &  0.13&  1.79 &  0.05 \\
CRT101665008         &4706 &   36&  2.74 &  0.10&  1.66 &  0.06 \\
CRT101748322         &4736 &   31&  2.60 &  0.07&  1.60 &  0.04 \\
KIC10525475          &4764 &   16&  2.45 &  0.06&  1.74 &  0.03 \\
KIC11394905          &4854 &   47&  2.34 &  0.16&  1.69 &  0.05 \\
KIC11445818 \tablenotemark{a}      &4767 & \nodata&  2.47 & \nodata&  1.81 & \nodata \\
KIC11823838          &4892 &   37&  2.40 &  0.13&  1.71 &  0.05 \\
KIC3455760           &4699 &   32&  2.66 &  0.07&  1.77 &  0.04 \\
KIC3833399           &4677 &   36&  2.36 &  0.14&  1.95 &  0.06 \\
KIC4143460           &4801 &   28&  2.44 &  0.09&  1.70 &  0.04 \\
KIC4350501           &4864 &   33&  2.99 &  0.09&  1.50 &  0.06 \\
KIC5512910           &4854 &   39&  2.33 &  0.13&  1.73 &  0.04 \\
KIC9269081           &4752 &   34&  2.25 &  0.09&  1.83 &  0.04 \\
KIC9821622           &4807 &   48&  2.69 &  0.12&  1.52 &  0.06 \\
  \enddata 
\tablenotetext{a}{Reference star.}
\end{deluxetable*}

Once the stellar parameters have been fixed, we derive abundance of each species for a star relative to the abundance for the reference star.
All the relative abundances are converted to the absolute scale using the abundance of the reference star, KIC~11445818, which is directly calculated using the line data listed in Table \ref{tablelinelist}.
Table \ref{tableab} lists the absolute abundances with uncertainties.
There are two sources of uncertainties in the derived abundance:
one is originated from measurements of equivalent widths and/or modeling of absorption lines, and the other is from uncertainties in stellar parameters.
The former appears as the line-to-line scatter ($\sigma$) in derived abundances and contribute to the total error as $\sigma/\sqrt{N}$, where $N$ is the number of lines for the species used in the analysis.
For the species that have less than three detectable lines and smaller line-to-line scatter than neutral iron lines, we adopt $\sigma$ of iron abundance from neutral iron lines.
The effect of uncertainties in stellar parameters is estimated by recalculating the abundances with the stellar parameters shifted by their uncertainties.
In this process, we take the correlation between parameters into account.
Since $T_{\rm eff}$ and $\log g$ are particularly degenerate in the process of stellar parameter determination, the covariance is sometimes important (Appendix).

\begin{deluxetable*}{llrrrr}
  \tablecaption{Abundances \label{tableab}}
  \tablehead{ 
  \colhead{Object} & \colhead{Species} & \dcolhead{[\mathrm{X/H}]} & \dcolhead{\sigma([\mathrm{X/H}])} & \dcolhead{[\mathrm{X/Fe}]} & \dcolhead{\sigma([\mathrm{X/Fe}])} 
 }
 \startdata
 2M0001+2415          & FeI   & -0.49 &  0.02 & \nodata & \nodata \\ 
 2M0001+2415          & FeII  & -0.47 &  0.04 & \nodata & \nodata \\ 
 2M0001+2415          & NaI   & -0.17 &  0.04 &  0.33 &  0.04 \\ 
 2M0001+2415          & MgI   & -0.23 &  0.04 &  0.26 &  0.04 \\ 
 2M0001+2415          & AlI   & -0.03 &  0.03 &  0.46 &  0.03 \\ 
  \enddata 
  \tablenotetext{}{Portion of the table is shown.}
  \tablenotetext{}{KIC11445818 is the reference star.}
\end{deluxetable*}

\section{Results\label{results}}
\subsection{Comparison to previous studies}
\subsubsection{Asteroseismology and parallax}
\begin{figure}
  \plotone{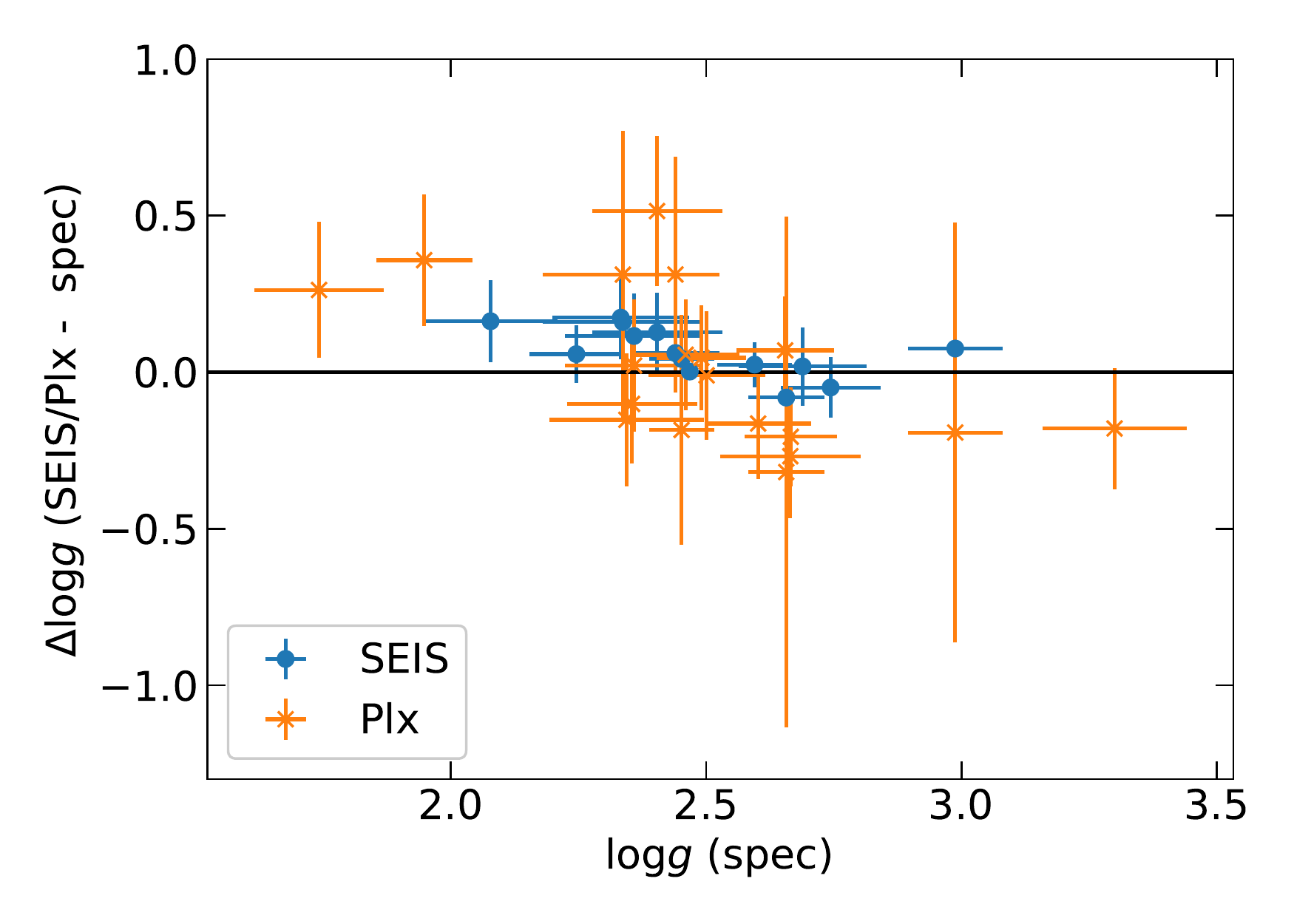}
  \caption{Comparison of spectroscopic $\log g$ with other methods. Averages and standard deviations of the differences are $0.06\pm 0.08$ for seismic $\log g$ and $0.00\pm 0.21$ for parallax $\log g$. \label{figloggcomp}}
\end{figure}

Figure \ref{figloggcomp} compares spectroscopic $\log g$ with those from asteroseismology and parallax.

All young $\alpha$-rich stars have asteroseismic information, which enables us to constrain $\log g$ with the precision of $\sim 0.01\,\mathrm{dex}$.
The $\log g$ from asteroseismology is calculated using the following relation:
\begin{equation}
g\propto \nu_{\rm max}T_{\rm eff}^{0.5}.
\end{equation}
Since we use KIC~11445818 as reference in the spectroscopic analysis, we derive asteroseismic $\log g$ in the scale of this star as 
\begin{equation}
\log g = 2.47 + \log(\nu_{\rm max}/37.05\,\mathrm{\mu Hz}) + 0.5\log(T_{\rm eff}/4767\,\mathrm{K}).
\end{equation}

While the comparison stars do not have asteroseismic information, twelve of them and seven of young $\alpha$-rich stars have parallax measurements, which also provides independent $\log g$ as
\begin{equation}
g = M/R^2 \propto M T_{\rm eff}^4 / L,
\end{equation}
where $M$ is the mass of the star, $R$ is the radius, and $L$ is the luminosity.
Luminosity is estimated from Gaia DR1/TGAS parallax \citep{GaiaCollaboration2016,GaiaCollaboration2016a}, $J,\,H,$ and $K_s$ magnitudes from 2MASS \citep{Cutri2003}, synthetic bolometric correction \citep{Casagrande2014}, and extinction \citep{Green2015}.
For the comparison stars, we adopt mass estimates of \citet{Feuillet2016}.
For the young $\alpha$-rich stars we adopt mass estimates from the scaling relation of asteroseismology as
\begin{equation}
M \propto \nu_{\rm max}^{-3} \Delta\nu ^4 T_{\rm eff}^{-1.5}.
\end{equation}

\added{The overall agreement is fairy good regardless of the choice of the method.}
Spectroscopic and parallax-based $\log g$ do not show any significant offsets (Figure \ref{figapogeecomp}), although the latter has larger error.
On the other hand, asteroseismic $\log g$ gives a slightly higher value compared to the spectroscopic method by $0.06\,\mathrm{dex}$ in average.  

\subsubsection{APOGEE}
\begin{figure*}
  \plotone{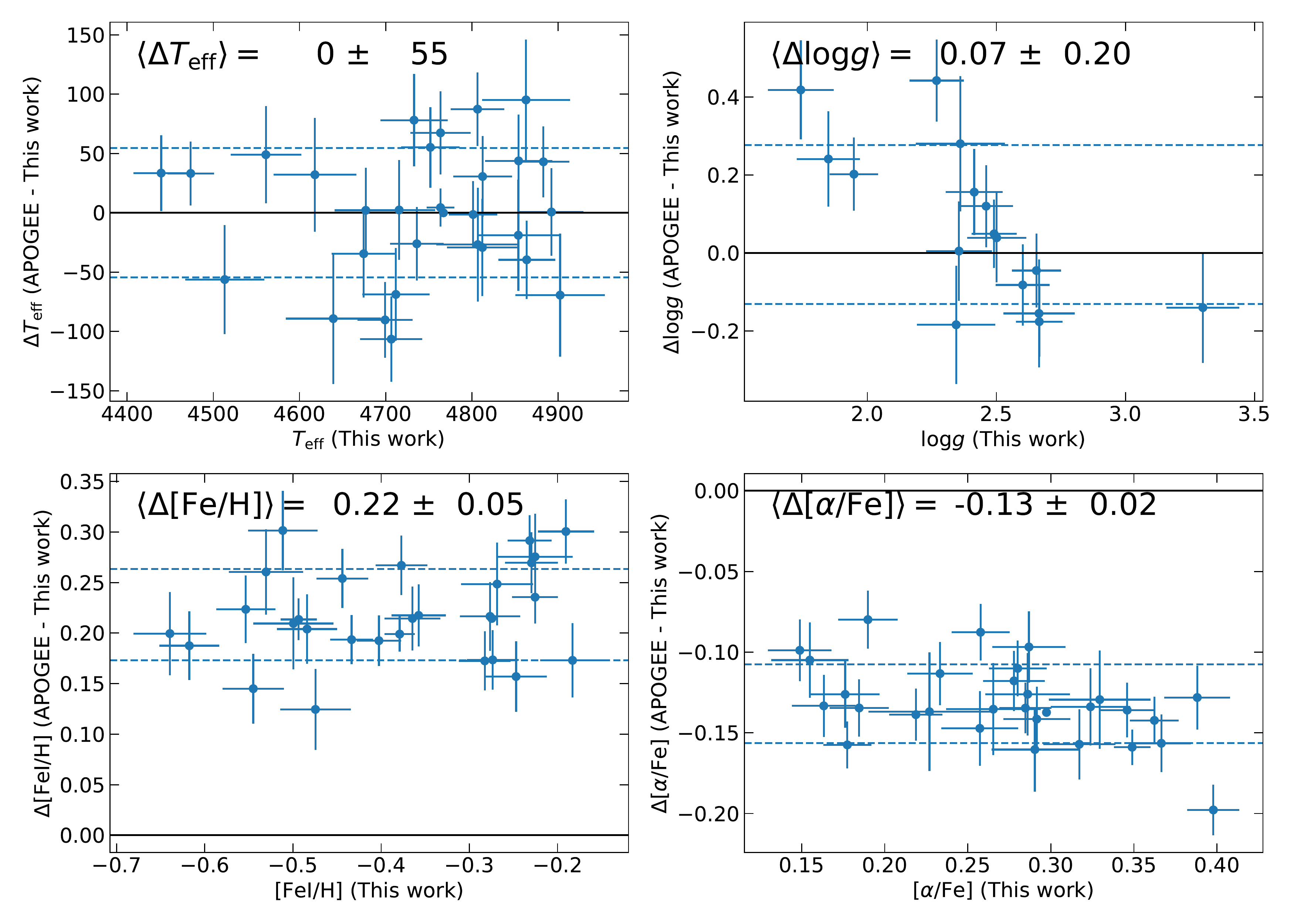}
  \caption{Comparison of stellar parameters with the APOGEE DR12 catalog. 
The average difference and standard deviation are shown in the top left corner of each panel.
Horizontal dashed lines show $1\sigma$ ranges of the difference. 
Note that error bars do not include the contribution of uncertainties in APOGEE.
\label{figapogeecomp}}
\end{figure*}
All of our stars are originally selected based on calibrated abundances of the APOGEE DR12 catalog. 
Here we compare $T_{\rm eff}$, $\log g$, [{Fe}/{H}], and [{$\alpha$}/{Fe}] with the values in \citet{Chiappini2015}, \citet{Martig2015a}, and \citet{Feuillet2016}.
The results are shown in Figure \ref{figapogeecomp}. 

Since we adopt $T_{\rm eff}$ of the reference star obtained from the APOGEE catalog, we expect good agreement on average in this parameter. 
In fact, there is no significant offset in $T_{\rm eff}$.
The amplitude of the star-to-star scatter is as small as expected from uncertainties of this work and APOGEE results. 

The other parameters show offsets with small dispersion.
We now briefly address the $0.22\pm 0.05\,\mathrm{dex}$ (standard deviation) offset in metallicity between our study and APOGEE results.
This offset could be coupled with the estimate of $\log g$.
Since we set the $\log g$ of the reference star from asteroseismology, we also compare our results with \citet{Hawkins2016}, who re-analysed APOGEE spectra of stars in the \textit{Kepler} field by utilizing asteroseismic $\log g$ constraints.
The metallicity offset is smaller ($0.08\,\mathrm{dex}$) between our results and \citet{Hawkins2016} work. 
The reason for the smaller offset is, however, unclear since they did not find systematic offset between their work and the calibrated APOGEE results in their whole sample.
We also tried another 1D/LTE spectral synthesis code that is used and described in, e.g., \citet{Aoki2009}, and confirmed that the results are unchanged.

The [{$\alpha$}/{Fe}] offset of $0.13\,\mathrm{dex}$ is also large compared to measurement errors.
This large offset is mainly due to our higher Si abundances than APOGEE results (see next subsection). 

\subsection{$\alpha$-elements}
\begin{figure*}
  \plotone{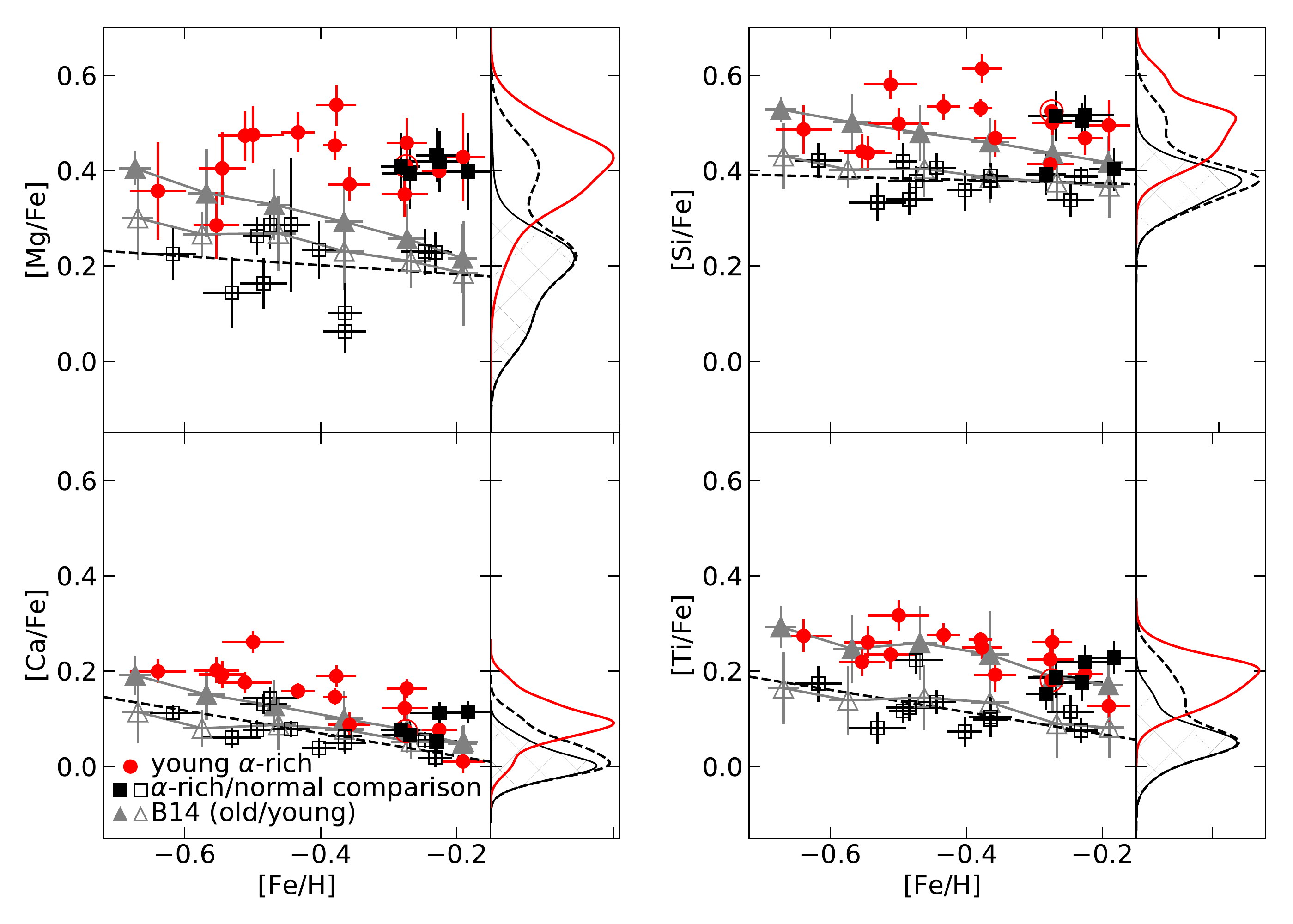}
  \caption{
Abundances of $\alpha$-elements as a function of metallicity for young $\alpha$-rich stars, our comparison stars, and \citet{Bensby2014} sample (B14).
The reference star is circled and shown without error bars. 
After correcting for the Galactic chemical evolution (black dashed line), we show the distribution of [{X}/{Fe}] of stars for each element and for each population (red solid; young $\alpha$-rich stars; black solid: $\alpha$-normal comparison stars; black dashed: whole comparison stars).
See text for details.
\label{figafe}}
\end{figure*}
\begin{figure}
  \plotone{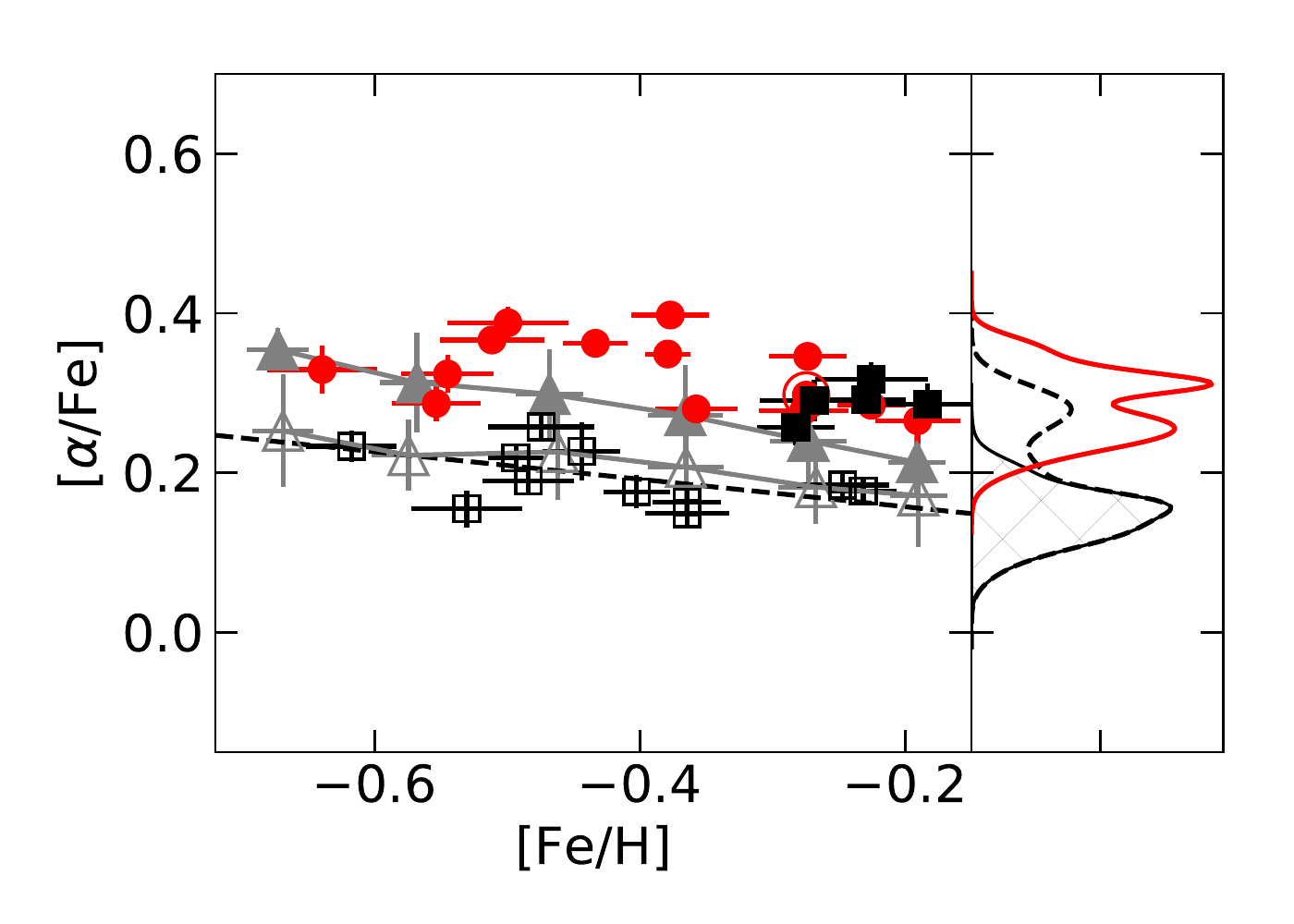}
  \caption{
The $[\mathrm{\alpha/Fe}]$ (mean of [{Mg}/{Fe}], [{Si}/{Fe}], [{Ca}/{Fe}], and [{Ti}/{Fe}]) as a function of metallicity.
Symbols are the same as in Figure \ref{figafe}.
\label{figalpha}}
\end{figure}

The abundances of $\alpha$-elements are shown in Figure \ref{figafe} and \ref{figalpha}, with the results from \citet{Bensby2014}.
For the \citet{Bensby2014} sample, we define old stars as those older than $9\,\mathrm{Gyr}$ and young stars as those younger than $7\,\mathrm{Gyr}$.
Since our sample and theirs differ in spectral types, the absolute scale could differ.
To take the advantages of differential analysis, the [{X/Fe}] of \citet{Bensby2014} is shifted by comparing the median abundance of $\alpha$-normal stars (see below) in our comparison sample and young stars in \citet{Bensby2014}.
The corrections are 0.06, 0.32, -0.02, and 0.03 for [{Mg}/{Fe}], [{Si}/{Fe}], [{Ca}/{Fe}], and [{Ti}/{Fe}], respectively.
The \citet{Bensby2014} sample is binned and median values for each bin are plotted.

We show the distributions of [{X}/{Fe}] after removing the effect of the Galactic chemical evolution.
The effect is determined from $\alpha$-normal comparison stars, using orthogonal distance regression. 
In Figure \ref{figafe} and subsequent similar plots, the zero point of the histogram is set to the [{X}/{Fe}] value of the chemical evolution at $[\mathrm{Fe/H}]=-0.15$ (metal-rich end of the left panel).

Overall $\alpha$-enhancements of young $\alpha$-rich stars compared to nearby red giants are clearly confirmed from Figure \ref{figalpha}.
We note that five stars in the comparison sample seem to have high [{$\alpha$}/{Fe}] for their metallicity.
These five stars are considered as $\alpha$-rich and treated separately from the rest of $\alpha$-normal comparison stars. 

Trends with metallicity among the individual $\alpha$-elements are similar between our $\alpha$-normal stars and young stars in \citet{Bensby2014}: 
[{$\alpha$}/{Fe}] decreases as metallicity increases.
The [{$\alpha$}/{Fe}] of old stars in \citet{Bensby2014} are also decreasing with metallicity, but are systematically higher than young stars.
Young $\alpha$-rich stars and $\alpha$-rich comparison stars generally follow this trend of old stars.

The $\alpha$-element abundances of young $\alpha$-rich stars obtained in this study are consistent with the results of \citet{Martig2015a} and \citet{Chiappini2015}. 
They have typical abundance pattern of old thick disk stars.

\subsection{Neutron-capture elements}
\begin{figure*}
  \plotone{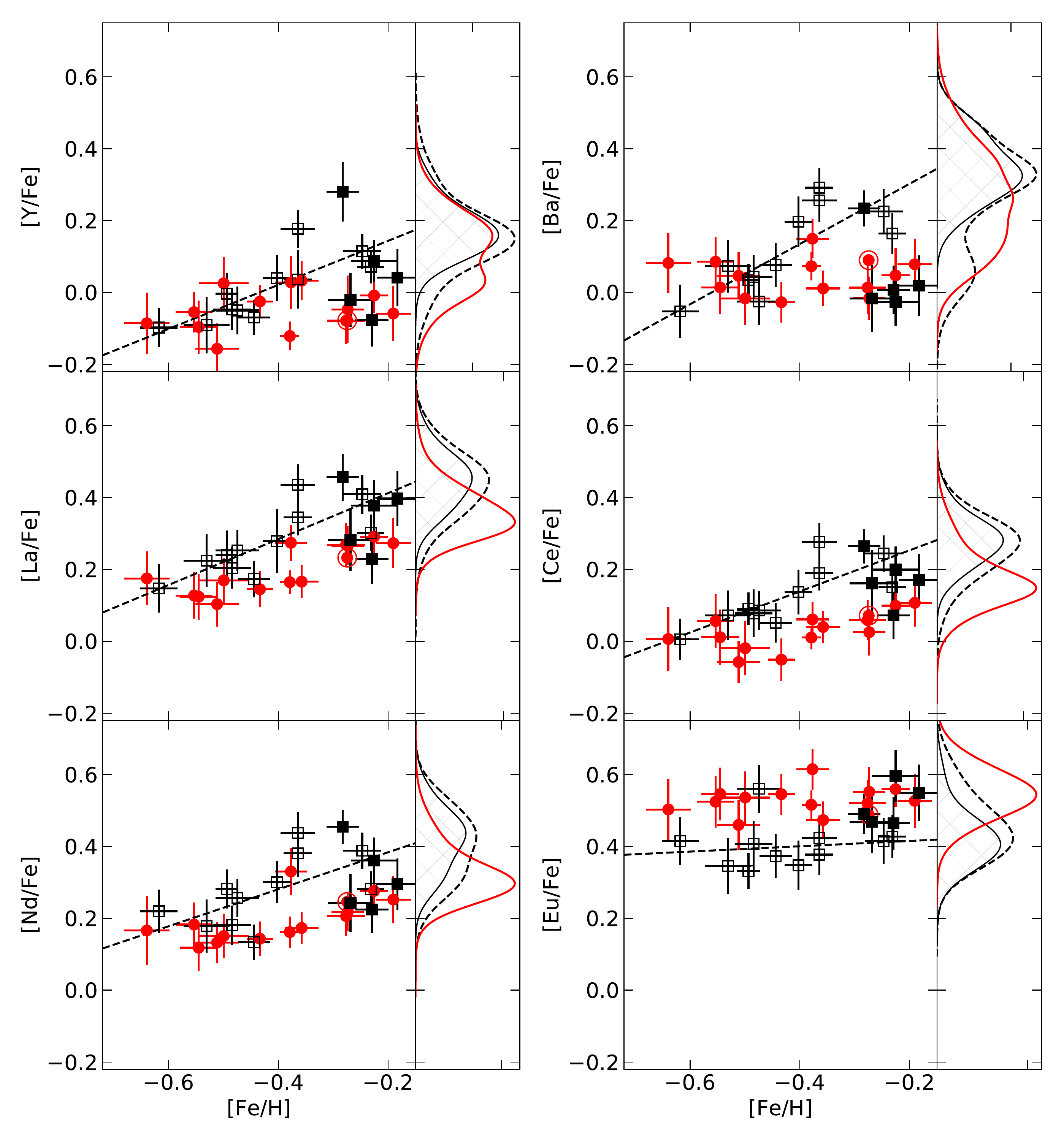}
  \caption{Abundances of neutron-capture elements. Symbols are the same as in Figure \ref{figafe}.\label{fignfe}}
\end{figure*}

\begin{figure}
  \plotone{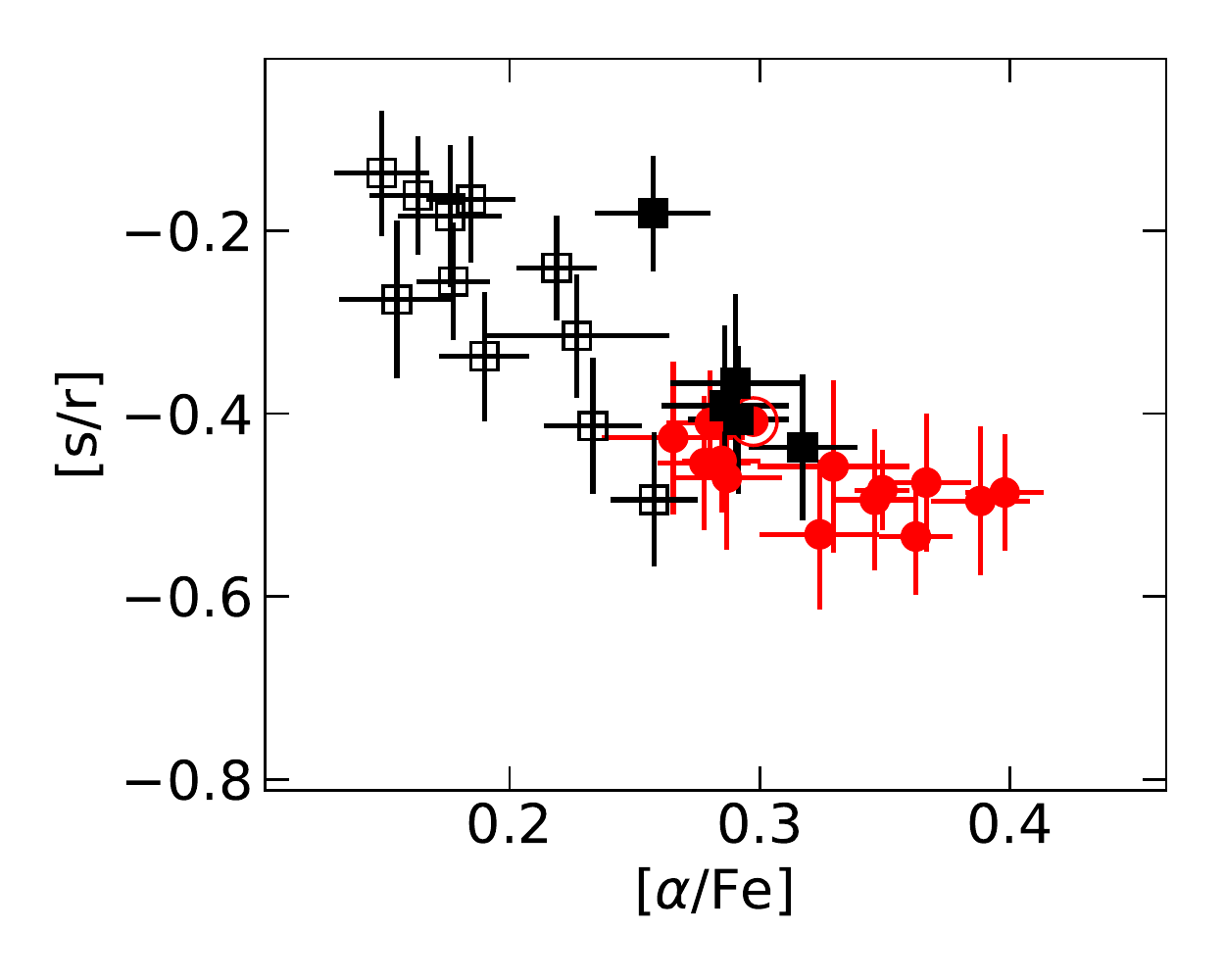}
  \caption{The [$s/r$] ratio as a function of [{$\alpha$}/{Fe}]. Symbols are the same as in Figure \ref{figafe}.\label{figsralpha}}
\end{figure}

While abundances of neutron-capture elements from APOGEE spectra of the young $\alpha$-rich stars are not yet available, our new measurements from optical spectra shed new light on the origin of these objects.
The $s$-process elements are produced mostly in the interior of low- to intermediate-mass stars \citep{Karakas2016} with a longer time-scale of enrichment than $\alpha$-elements.
The $r$-process elements are considered to be efficiently synthesized by neutron star mergers \citep[e.g., ][]{Wanajo2014,Drout2017}.
Though the time scale of the enrichment is still uncertain, the early rise of $r$-process abundances in metal-poor stars indicates a shorter time-scale compared to that by type~Ia supernovae or $s$-process (\citealt{Argast2004}, see also \citealt{Ishimaru2015,Hirai2015,Hotokezaka2018})\added{, whereas some sort of supernovae are proposed to be another source \citep{Nishimura2017}}.
Recent high-precision abundance analysis of solar twins illustrates this timescale difference of abundance trends of neutron-capture elements \citep{Spina2018}.

We determine abundances of Y II, Ba II, La II, Ce II, Nd II, and Eu II, among which the $r$-process has a dominant contribution only to Eu \citep{Sneden2008}.
Their abundance distributions are shown in Figure \ref{fignfe}.
Although \citet{Battistini2016} derived abundances of neutron-capture elements for the sample of \citet{Bensby2014}, there are not many measurements of neutron-capture element abundances at low metallicity and the measured abundances show large scatter.
Therefore, we do not include their results in figures.

The abundance ratio [{X}/{Fe}] for $s$-process elements increases with metallicity in $\alpha$-normal stars (Figure \ref{fignfe}).
On the other hand, $s$-process abundances of young $\alpha$-rich stars and $\alpha$-rich comparison stars are almost flat, which results in lower $s$-process abundances at high metallicity. 
This near constancy in $s$-process elements supports a short time-scale for star formation as suggested from $\alpha$-elements and Fe.

In contrast to the $s$-process elements, the Eu abundance is higher in young $\alpha$-rich stars as well as $\alpha$-rich comparison stars than in $\alpha$-normal stars, i.e., similar feature as $\alpha$-elements.
This result is naturally understood from short timescale of $r$-process enrichments, which has been observationally shown.

To see the relative contribution of $s$- and $r$-processes, we also investigate [$s/r$] as a function of [{$\alpha$}/{Fe}] (Figure \ref{figsralpha}). 
Here [$s$/{Fe}] is the average of [{X}/{Fe}] for Y~II, Ba~II, La~II, and Ce~II, and $[s/r]=[s/\mathrm{Fe}]-[\mathrm{Eu/Fe}]$.
Nucleosynthesis events with short time-scales such as $r$-process production events or type~II supernovae mean that the starting point of chemical evolution are high [{$\alpha$}/{Fe}] and low [$s/r$] ratios.
The later contribution from low- to intermediate-mass stars produce more Fe and $s$-process elements than the earlier events and move stars towards the upper left in Figure \ref{figsralpha}.
On the other hand, mass transfer from AGB companions in binary systems should lead to high [$s/r$] ratio with no change of [$\alpha$/{Fe}]. Such objects are not found in Figure \ref{figsralpha}.

Young $\alpha$-rich stars and nearby bright giants mostly follow the same trend in Figure \ref{figsralpha}.
Importantly, none of the young $\alpha$-rich stars are $s$-process enhanced.
One of the key results of this work is that young $\alpha$-rich stars are not chemically peculiar in neutron-capture elements.

\subsection{Other elements (Li, and Na through Cu)}
\begin{figure*}
  \plotone{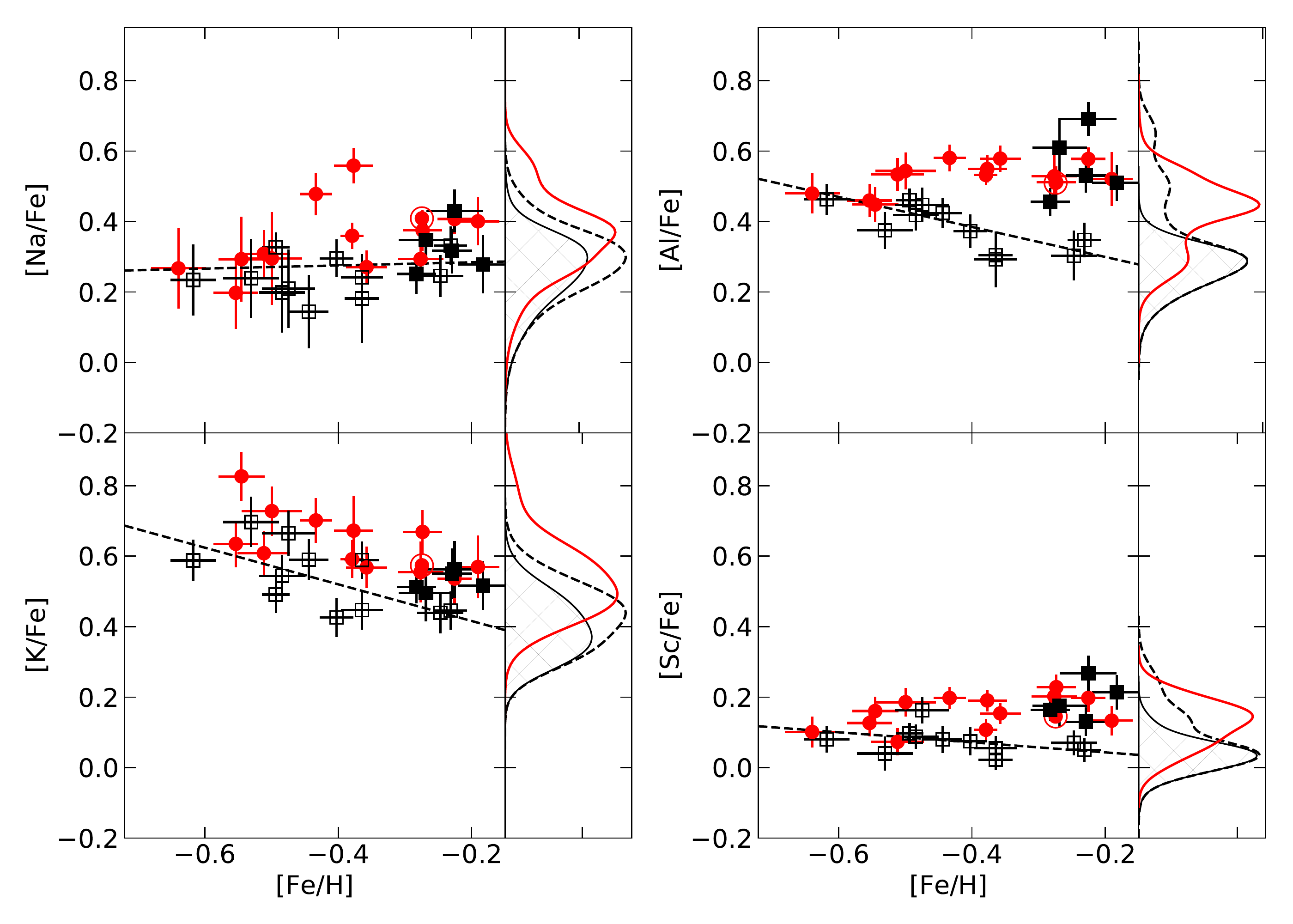}
  \caption{Abundances of Na~I, Al~I, K~I, and Sc~I. Symbols are the same as in Figure \ref{figafe}.\label{figoddz}}
\end{figure*}

\begin{figure*}
  \plotone{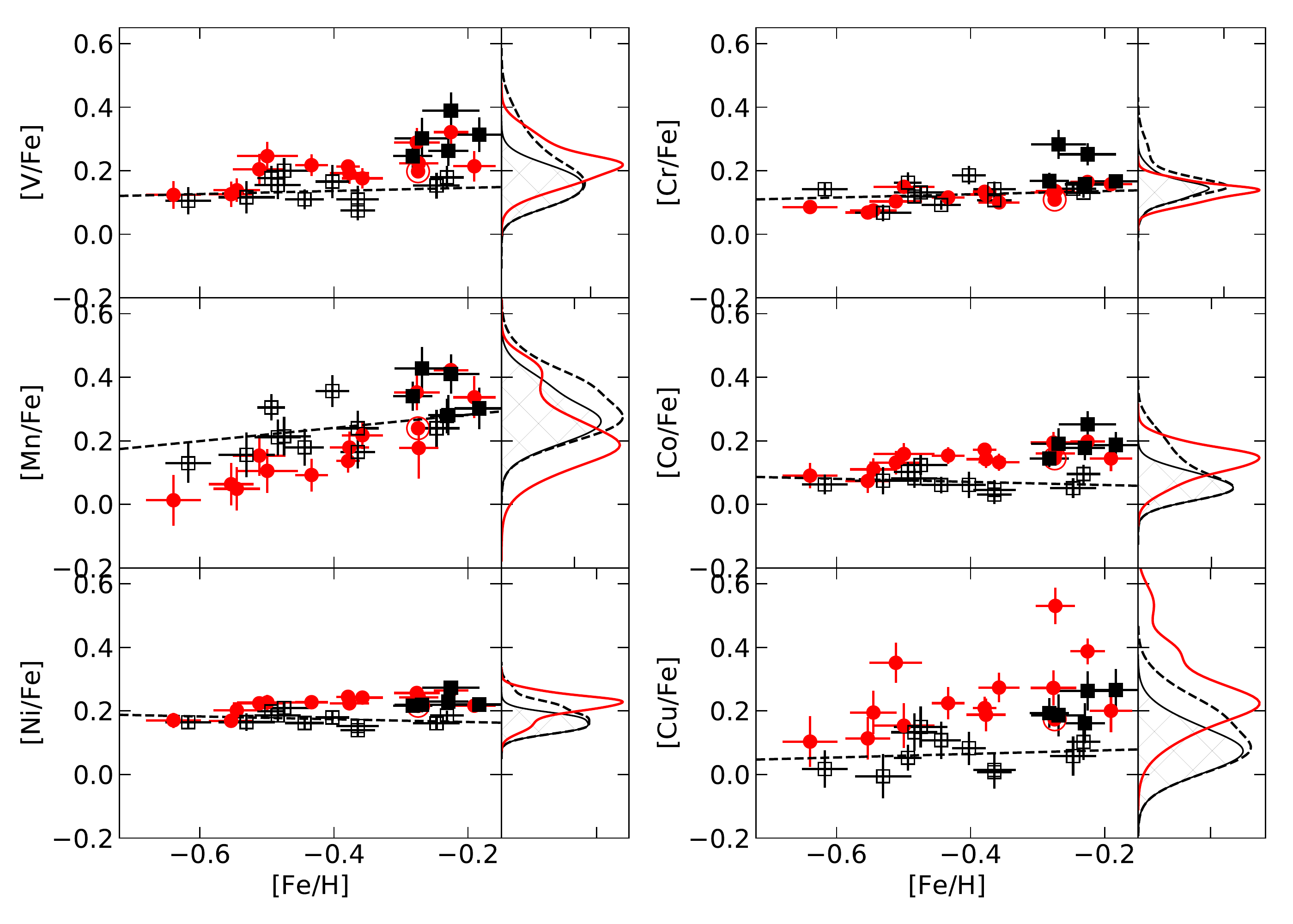}
  \caption{Abundances of V~I, Cr~I, Mn~I, Co~I, Ni~I, and Cu~I. Symbols are the same as in Figure \ref{figafe}.\label{figironpeak}}
\end{figure*}

Abundances of other elements from Na to Cu are shown in Figure \ref{figoddz} and \ref{figironpeak}.
Some of those elements have also been measured in APOGEE and presented in \citet{Chiappini2015}.

The overall distribution of the abundances of young $\alpha$-rich stars is similar to the $\alpha$-rich comparison sample.
This is expected from the similarities in $\alpha$-elements and neutron-capture elements.

There are some young $\alpha$-rich stars that seem to show mild enhancement in Na or Cu.
Since the two absorption lines used for the analysis of Na at $6154\,\mathrm{\AA}$ and $6161\,\mathrm{\AA}$ are free from blending of other features, the measurements of equivalent widths are quite robust.
By contrast, the only absorption line of Cu in the analysis is located at $5782\,\mathrm{\AA}$, a relatively crowded region in which the continuum placement is difficult.

The Na abundances of the two stars with high [{Na}/{Fe}] in this study (KIC~4143460 and KIC~9269081) are also high in \citet{Hawkins2016}. 
The Na enhancements might be related to their high-mass \citep[e.g.,][]{Luck1994}, especially for the case of KIC~9269081.  

KIC~9821622 was reported to be Li-rich in \citet{Jofre2015}, which is later confirmed by \citet{Yong2016}.
The fraction of Li-rich objects among red giants is estimated to be $\sim 1\,\%$, and abundances of other elements have been shown to be indistinguishable from normal giants \citep{Takeda2017}.
One of the scenarios proposed to explain the Li-excess is engulfment of a brown dwarf or planet, that could accompany mass increase.
If young $\alpha$-rich stars have obtained mass through such engulfment, we would expect high fraction of Li-rich objects.

Among 14 objects, only KIC~9821622 shows clear Li-enhancement, which can be seen at $\sim 6708\,\mathrm{\AA}$.
The line is not detectable for the others. 
\added{We estimate the fraction of Li-rich stars among young $\alpha$-rich stars as $0.07^{+0.13}_{-0.02}\%$ ($1\sigma$) and $0.07^{+0.29}_{-0.06}$ ($2\sigma$) from binominal distribution.}
Due to the limited size of the sample, we cannot conclude if the fraction of Li-enhanced objects is high.

\subsection{Radial velocities and line widths}

Radial velocity variation is a sign of the existence of a companion.
If mass transfer has played an important role in the formation of young $\alpha$-rich stars, a significant fraction of them should exhibit radial velocity variations \citep{Izzard2018,Jofre2016}. 

Radial velocity monitoring of 13 young $\alpha$-rich stars has been carried out by \citet{Jofre2016}.
They report that six stars have high probability ($P>0.68$) of radial velocity variation.
They found that the fraction of stars with radial velocity variation is higher for young $\alpha$-rich stars than for their comparison stars, though this difference is not significant at $2\sigma$ level. 

We found that seven out of the 14 young $\alpha$-rich stars in our study have radial velocity variation larger than $1\,\mathrm{km\,s^{-1}}$.
By contrast, only two out of the 16 comparison stars show the variation. 
This difference is significant at the $2\sigma$ level according to Fisher's exact test ($p=0.046$).

The $1\,\mathrm{km\,s^{-1}}$ criterion is rather conservative (see also Section \ref{obs}). 
We also carry out the same test adopting a less conservative $0.7\,\mathrm{km\,s^{-1}}$ as the criterion and obtain $p=0.057$.
We conclude that the choice of the criterion does not affect our conclusion.

All of our eleven young $\alpha$-rich stars in the \textit{Kepler} field have been studied by \citet{Jofre2016}.
While KIC~9821622 and KIC~11394905 are not regarded as members of binary systems in this study, they have high probability of radial velocity variation in \citet{Jofre2016}. 
On the other hand, KIC~3833399 shows radial velocity change from APOGEE to our observation, but is not identified as a binary star in \citet{Jofre2016}.
If we combine our results and \citet{Jofre2016}, \replaced{more than}{at least} eight out of the 13 young $\alpha$-rich stars in the \textit{Kepler} field show radial velocity variation larger than $1\,\mathrm{km\,s^{-1}}$.
Thanks to the increase of the baseline of radial velocity monitoring, we increase the likelihood that young $\alpha$-rich stars belong to binary systems, which supports the mass transfer scenario.

Stellar mergers and mass transfer could cause excess of rotation velocity as a result of angular momentum transfer.
We investigate the median of FWHM of iron lines with $-5.5<\log (EW/\lambda)<-5.0$ to search for such signatures.
The average of the medians of FWHM for the comparison sample is $ 7.9 \pm  0.4\,\mathrm{km\,s^{-1}}$ (standard deviation), while that for young $\alpha$-rich stars is $8.0\pm0.4\,\mathrm{km\,s^{-1}}$. 
The largest values are $9.1\,\mathrm{km\,s^{-1}}$ among the comparisons and $8.8\,\mathrm{km\,s^{-1}}$ among the young $\alpha$-rich stars.
We conclude that there is no signature of broad absorption features in young $\alpha$-rich stars.

\section{Discussion\label{discussion}}

Our findings in the Section \ref{results} are summarised as follows: i) young $\alpha$-rich stars have similar abundance trends to old $\alpha$-rich stars. ii) they exhibit a high binary frequency. 
These results point to mass transfer from a companion as the dominant formation channel of young $\alpha$-rich stars \citep{Izzard2018}.
In particular, the lack of $s$-process enhancement suggests that the binary companions were not AGB stars.

The lack of $s$-process enhancement is well explained by the scenario of mass transfer origin. 
Since stars that have accreted significant amount of mass cannot survive long due to their large masses, binary interaction should have occurred recently for such stars to be observed at present.
Then, the companion cannot be too massive since we assume it was co-eval and survived until the recent interaction.

\citet{Chiappini2015} and \citet{Martig2015a} derived $\lesssim 5\,\mathrm{Gyr}$ upper limits on age of young $\alpha$-rich stars based on the assumption that they were born with the mass currently estimated from asteroseismology. 
This means that the mass transfer events occurred within the past $5\,\mathrm{Gyr}$.
On the other hand, the typical age of $\alpha$-rich thick disk stars is $>8\,\mathrm{Gyr}$.
Accordingly, the companion should have lived for at least $3\,\mathrm{Gyr}$, which corresponds to the lifetime of $\sim 1.5\,\mathrm{M_{\odot}}$ solar metallicity stars.
This estimate is conservative, so we expect that most of the companions were less massive than $1.5\,\mathrm{M_{\odot}}$.
Since stars with $M\lesssim 1.3\,\mathrm{M_{\odot}}$ do not produce substantial amounts of $s$-process elements \citep{Karakas2016}, we cannot expect many of the companions have produced significant amount of $s$-process elements.

Not all stars show radial velocity variation, which might indicate that some of the stars formed through stellar merger. 
In order to constrain the formation channel, we need to keep monitoring the radial velocities for these stars.
Additional radial velocity measurements for the stars with radial velocity variation is also important to constrain the nature of the companion through its mass determination.

\acknowledgements
We thank Dr. Paula Jofr\'e, Dr. Luca Casagrande, and Prof. Martin Asplund for fruitful discussion, \added{and the referee for his/her comments}. 
This work is financially supported by the course-by-course education program of SOKENDAI.
The data presented herein were obtained at the W.M. Keck Observatory, which is operated as a scientific partnership among the California Institute of Technology, the University of California and the National Aeronautics and Space Administration. The Observatory was made possible by the generous financial support of the W.M. Keck Foundation. The authors wish to recognize and acknowledge the very significant cultural role and reverence that the summit of Maunakea has always had within the indigenous Hawaiian community. We are most fortunate to have the opportunity to conduct observations from this mountain. 

\software{NumPy \citep{numpy},
          SciPy \citep{scipy},
          Astropy \citep{Astropy},
          Matplotlib \citep{Matplotlib},
          Pandas \citep{pandas},
          Corner \citep{corner}
         }
\appendix
\section{Stellar Parameter Determination}
Stellar parameters ($T_{\rm eff}, \log g, v_t$) are determined by incorporating a Markov chain Monte Carlo method (MCMC) to q$^2$.
In exchange for computational costs, use of MCMC has advantages that correlations between stellar parameters can be estimated and that an exact convergence required in the original q$^2$ is not needed in calculations.

As described in the main text, the determination of stellar parameters is based on the analysis of iron lines.
In what follows, we use $\Delta A(\mathrm{Fe})_i$ as the relative iron abundance derived for the line $i$.
Following the traditional determination method, we evaluate three quantities: correlation between $\Delta A(\mathrm{Fe})_i$ and the reduced equivalent width $REW_i=\log(EW_i/\lambda_i)$ for neutral iron lines ($r_{(EW,A)}$), correlation between $\Delta A(\mathrm{Fe})_i$ and $\chi_i$ for neutral iron lines ($r_{(\chi, A)}$), and the difference of averages of $\Delta A(\mathrm{Fe})$ of neutral and singly-ionized lines $dA=\langle \Delta A(\mathrm{Fe})_i \rangle_{\rm Fe I} - \langle \Delta A(\mathrm{Fe})_i \rangle_{\rm Fe II}$\footnote{$r_{(X,Y)}$ denotes the Pearson's correlation coefficient between variables $X$ and $Y$.}.
We assume that when we use the appropriate set of stellar parameters, each $\Delta A(\mathrm{Fe})_i$ is distributed randomly around the true relative abundance. 
In other words, the basic assumption is that when the uncertainties of measured equivalent widths are infinitesimal and when the proper stellar parameters are used in the analysis of a star, relative iron abundances derived for individual iron lines $\Delta A(\mathrm{Fe})_i$ would be the same.

The likelihood can be decomposed as 
\begin{equation}
\mathcal{L}(r_{(EW,A)},\,r_{(\chi ,A)},\, dA| T_{\rm eff},\,\log g,\,v_t) = f(r_{(EW, A)},\,r_{(\chi, A)})\times f(dA).
\end{equation}
Note that $r_{(EW, A)}$ and $r_{(\chi ,A)}$ are not independent because both involve $\Delta A(\mathrm{Fe})_i$.
Therefore, the distribution of $(r_{(EW,A)},r_{(\chi ,A)})$ should be investigated under the presence of $r_{(EW,\chi)}$, and hence, 
\begin{eqnarray}
f(r_{(EW,A)},r_{(\chi, A)}) = & f(r_{(EW,A)},r_{(\chi, A)}|r_{(EW, \chi)}) \\
=& \frac{f(r_{(EW, \chi)}|r_{(EW,A)},r_{(\chi ,A)})f(r_{(EW,A)})f(r_{(\chi ,A)})}{f(r_{(EW, \chi)})},
\end{eqnarray}
where $f(r_{(EW,A)}),\,f(r_{(\chi ,A)}),$ and $f(r_{(EW, \chi)})$ are the probability distribution function of Pearson's correlation coefficient for an independent set of two variables, and $f(r_{(EW, \chi)}|r_{(EW,A)},r_{(\chi ,A)})$ is that of partial correlation coefficient.
The $f(dA)$ is calculated using Student's $t$-test described in the Section 14.2 of \citet{Press1992}.

We use flat priors with boundaries on $T_{\rm eff},\log g, v_t$.
The initial guess is adopted from the estimates by q$^2$, though the choice of the initial guess does not affect the final results.
The EnsembleSampler in \texttt{emcee} \citep{emcee} was used to sample the \added{posterior probability} distributions. 
The median of each parameter is adopted in the abundance analysis, and the half of the difference between 16th and 84th percentile values is adopted as the uncertainty.
Correlation coefficients between parameters are estimated and used in the error estimation of the abundances.

\begin{figure}
  \plotone{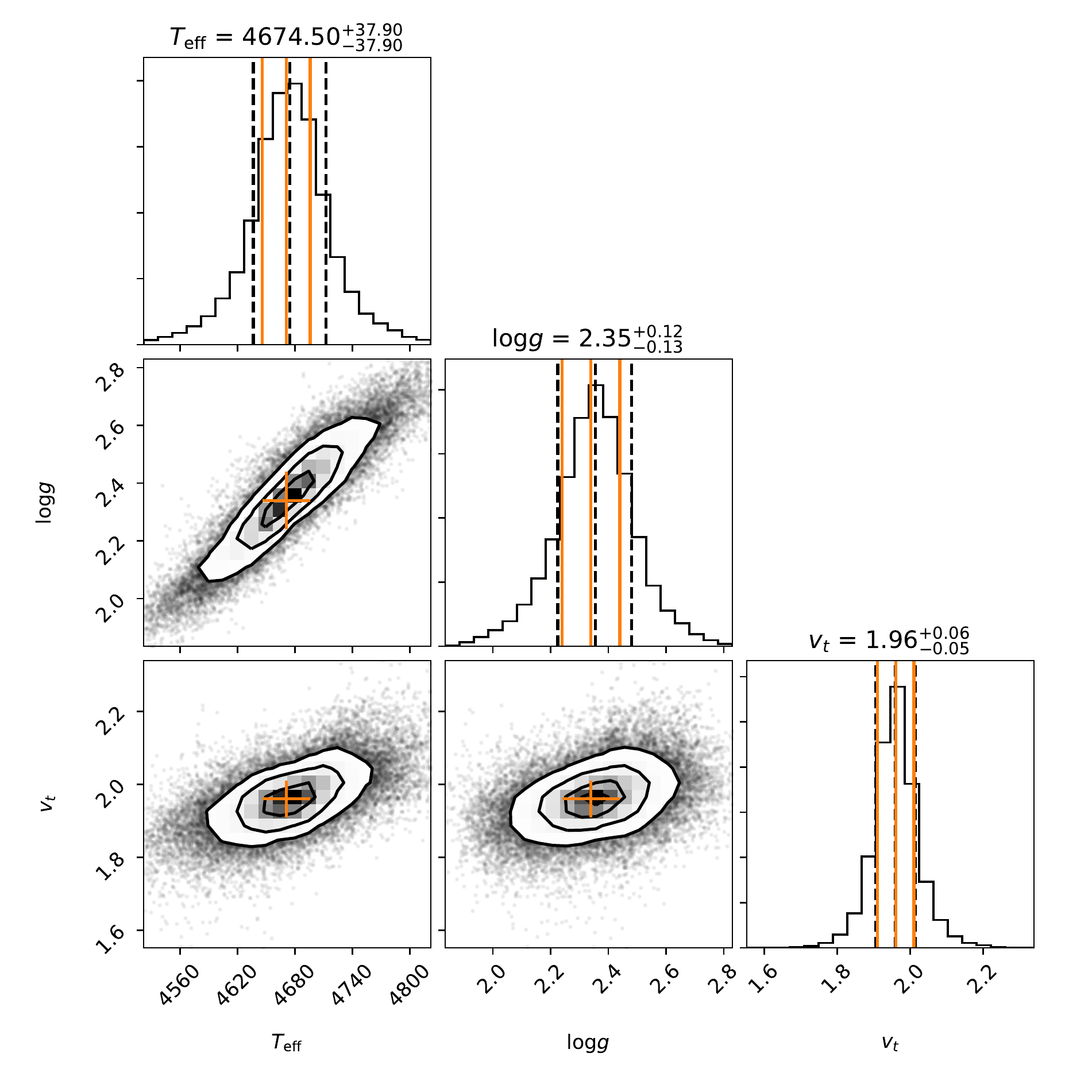}
  \caption{Corner plots to show the result of the MCMC stellar parameter determination. Black dashed lines in the histograms show 16th, 50th, and 84th percentile values. Orange plus signs in corner plots and solid lines in histograms show the parameters and $1\sigma$ uncertainties given by the original q$^2$\label{figmcmc_ex}.}
\end{figure}

In Figure \ref{figmcmc_ex}, we show an example of scatter plots for 2M0040+5927 with the result from q$^2$. 
It is clear that the two methods are basically consistent.
The effectiveness of the MCMC method is that we can estimate the correlations between parameters.
In principle, we can adopt a prior from asteroseismology or parallax measurements, which serves to resolve degeneracies. 

\bibliographystyle{aasjournal}

\begin{thebibliography}{}
\expandafter\ifx\csname natexlab\endcsname\relax\def\natexlab#1{#1}\fi
\providecommand{\url}[1]{\href{#1}{#1}}

\bibitem[{{Aerts} {et~al.}(2010){Aerts}, {Christensen-Dalsgaard}, \&
  {Kurtz}}]{Aerts2010}
{Aerts}, C., {Christensen-Dalsgaard}, J., \& {Kurtz}, D.~W. 2010,
  {Asteroseismology} (Springer, Dordrecht)

\bibitem[{{Aldenius} {et~al.}(2009){Aldenius}, {Lundberg}, \&
  {Blackwell-Whitehead}}]{Aldenius2009}
{Aldenius}, M., {Lundberg}, H., \& {Blackwell-Whitehead}, R. 2009, \aap, 502,
  989

\bibitem[{{Amarsi} {et~al.}(2016){Amarsi}, {Asplund}, {Collet}, \&
  {Leenaarts}}]{Amarsi2016}
{Amarsi}, A.~M., {Asplund}, M., {Collet}, R., \& {Leenaarts}, J. 2016, \mnras,
  455, 3735

\bibitem[{{Aoki} {et~al.}(2009){Aoki}, {Barklem}, {Beers}, {Christlieb},
  {Inoue}, {Garc{\'{\i}}a P{\'e}rez}, {Norris}, \& {Carollo}}]{Aoki2009}
{Aoki}, W., {Barklem}, P.~S., {Beers}, T.~C., {et~al.} 2009, \apj, 698, 1803

\bibitem[{{Argast} {et~al.}(2004){Argast}, {Samland}, {Thielemann}, \&
  {Qian}}]{Argast2004}
{Argast}, D., {Samland}, M., {Thielemann}, F.-K., \& {Qian}, Y.-Z. 2004, \aap,
  416, 997

\bibitem[{{Asplund} {et~al.}(2009){Asplund}, {Grevesse}, {Sauval}, \&
  {Scott}}]{Asplund2009}
{Asplund}, M., {Grevesse}, N., {Sauval}, A.~J., \& {Scott}, P. 2009, \araa, 47,
  481

\bibitem[{{Astropy Collaboration} {et~al.}(2013){Astropy Collaboration},
  {Robitaille}, {Tollerud}, {Greenfield}, {Droettboom}, {Bray}, {Aldcroft},
  {Davis}, {Ginsburg}, {Price-Whelan}, {Kerzendorf}, {Conley}, {Crighton},
  {Barbary}, {Muna}, {Ferguson}, {Grollier}, {Parikh}, {Nair}, {Unther},
  {Deil}, {Woillez}, {Conseil}, {Kramer}, {Turner}, {Singer}, {Fox}, {Weaver},
  {Zabalza}, {Edwards}, {Azalee Bostroem}, {Burke}, {Casey}, {Crawford},
  {Dencheva}, {Ely}, {Jenness}, {Labrie}, {Lim}, {Pierfederici}, {Pontzen},
  {Ptak}, {Refsdal}, {Servillat}, \& {Streicher}}]{Astropy}
{Astropy Collaboration}, {Robitaille}, T.~P., {Tollerud}, E.~J., {et~al.} 2013,
  \aap, 558, A33

\bibitem[{{Bard} {et~al.}(1991){Bard}, {Kock}, \& {Kock}}]{Bard1991}
{Bard}, A., {Kock}, A., \& {Kock}, M. 1991, \aap, 248, 315

\bibitem[{{Bard} \& {Kock}(1994)}]{Bard1994}
{Bard}, A., \& {Kock}, M. 1994, \aap, 282, 1014

\bibitem[{{Battistini} \& {Bensby}(2016)}]{Battistini2016}
{Battistini}, C., \& {Bensby}, T. 2016, \aap, 586, A49

\bibitem[{{Bensby} {et~al.}(2014){Bensby}, {Feltzing}, \& {Oey}}]{Bensby2014}
{Bensby}, T., {Feltzing}, S., \& {Oey}, M.~S. 2014, \aap, 562, A71

\bibitem[{{Blackwell} {et~al.}(1986){Blackwell}, {Booth}, {Haddock}, {Petford},
  \& {Leggett}}]{Blackwell1986}
{Blackwell}, D.~E., {Booth}, A.~J., {Haddock}, D.~J., {Petford}, A.~D., \&
  {Leggett}, S.~K. 1986, \mnras, 220, 549

\bibitem[{{Blackwell} {et~al.}(1979){Blackwell}, {Petford}, \&
  {Shallis}}]{Blackwell1979}
{Blackwell}, D.~E., {Petford}, A.~D., \& {Shallis}, M.~J. 1979, \mnras, 186,
  657

\bibitem[{{Blackwell} {et~al.}(1980){Blackwell}, {Petford}, {Shallis}, \&
  {Simmons}}]{Blackwell1980}
{Blackwell}, D.~E., {Petford}, A.~D., {Shallis}, M.~J., \& {Simmons}, G.~J.
  1980, \mnras, 191, 445

\bibitem[{{Blackwell} {et~al.}(1982{\natexlab{a}}){Blackwell}, {Petford},
  {Shallis}, \& {Simmons}}]{Blackwell1982}
---. 1982{\natexlab{a}}, \mnras, 199, 43

\bibitem[{{Blackwell} {et~al.}(1982{\natexlab{b}}){Blackwell}, {Petford}, \&
  {Simmons}}]{Blackwell1982a}
{Blackwell}, D.~E., {Petford}, A.~D., \& {Simmons}, G.~J. 1982{\natexlab{b}},
  \mnras, 201, 595

\bibitem[{{Booth} {et~al.}(1984){Booth}, {Blackwell}, {Petford}, \&
  {Shallis}}]{Booth1984}
{Booth}, A.~J., {Blackwell}, D.~E., {Petford}, A.~D., \& {Shallis}, M.~J. 1984,
  \mnras, 208, 147

\bibitem[{{Casagrande} \& {VandenBerg}(2014)}]{Casagrande2014}
{Casagrande}, L., \& {VandenBerg}, D.~A. 2014, \mnras, 444, 392

\bibitem[{{Castelli} \& {Kurucz}(2004)}]{Castelli2004}
{Castelli}, F., \& {Kurucz}, R.~L. 2004, ArXiv Astrophysics e-prints,
  astro-ph/0405087

\bibitem[{{Chaplin} {et~al.}(2011){Chaplin}, {Kjeldsen},
  {Christensen-Dalsgaard}, {Basu}, {Miglio}, {Appourchaux}, {Bedding},
  {Elsworth}, {Garc{\'{\i}}a}, {Gilliland}, {Girardi}, {Houdek}, {Karoff},
  {Kawaler}, {Metcalfe}, {Molenda-{\.Z}akowicz}, {Monteiro}, {Thompson},
  {Verner}, {Ballot}, {Bonanno}, {Brand{\~a}o}, {Broomhall}, {Bruntt},
  {Campante}, {Corsaro}, {Creevey}, {Do{\u g}an}, {Esch}, {Gai}, {Gaulme},
  {Hale}, {Handberg}, {Hekker}, {Huber}, {Jim{\'e}nez}, {Mathur}, {Mazumdar},
  {Mosser}, {New}, {Pinsonneault}, {Pricopi}, {Quirion}, {R{\'e}gulo},
  {Salabert}, {Serenelli}, {Silva Aguirre}, {Sousa}, {Stello}, {Stevens},
  {Suran}, {Uytterhoeven}, {White}, {Borucki}, {Brown}, {Jenkins}, {Kinemuchi},
  {Van Cleve}, \& {Klaus}}]{Chaplin2011}
{Chaplin}, W.~J., {Kjeldsen}, H., {Christensen-Dalsgaard}, J., {et~al.} 2011,
  Science, 332, 213

\bibitem[{{Chiappini} {et~al.}(2015){Chiappini}, {Anders}, {Rodrigues},
  {Miglio}, {Montalb{\'a}n}, {Mosser}, {Girardi}, {Valentini}, {Noels},
  {Morel}, {Minchev}, {Steinmetz}, {Santiago}, {Schultheis}, {Martig}, {da
  Costa}, {Maia}, {Allende Prieto}, {de Assis Peralta}, {Hekker},
  {Theme{\ss}l}, {Kallinger}, {Garc{\'{\i}}a}, {Mathur}, {Baudin}, {Beers},
  {Cunha}, {Harding}, {Holtzman}, {Majewski}, {M{\'e}sz{\'a}ros}, {Nidever},
  {Pan}, {Schiavon}, {Shetrone}, {Schneider}, \& {Stassun}}]{Chiappini2015}
{Chiappini}, C., {Anders}, F., {Rodrigues}, T.~S., {et~al.} 2015, \aap, 576,
  L12

\bibitem[{{Cutri} {et~al.}(2003){Cutri}, {Skrutskie}, {van Dyk}, {Beichman},
  {Carpenter}, {Chester}, {Cambresy}, {Evans}, {Fowler}, {Gizis}, {Howard},
  {Huchra}, {Jarrett}, {Kopan}, {Kirkpatrick}, {Light}, {Marsh}, {McCallon},
  {Schneider}, {Stiening}, {Sykes}, {Weinberg}, {Wheaton}, {Wheelock}, \&
  {Zacarias}}]{Cutri2003}
{Cutri}, R.~M., {Skrutskie}, M.~F., {van Dyk}, S., {et~al.} 2003, VizieR Online
  Data Catalog, 2246

\bibitem[{{Den Hartog} {et~al.}(2003){Den Hartog}, {Lawler}, {Sneden}, \&
  {Cowan}}]{DenHartog2003}
{Den Hartog}, E.~A., {Lawler}, J.~E., {Sneden}, C., \& {Cowan}, J.~J. 2003,
  \apjs, 148, 543

\bibitem[{{Den Hartog} {et~al.}(2011){Den Hartog}, {Lawler}, {Sobeck},
  {Sneden}, \& {Cowan}}]{DenHartog2011}
{Den Hartog}, E.~A., {Lawler}, J.~E., {Sobeck}, J.~S., {Sneden}, C., \&
  {Cowan}, J.~J. 2011, \apjs, 194, 35

\bibitem[{{Drout} {et~al.}(2017){Drout}, {Piro}, {Shappee}, {Kilpatrick}, {Simon}, {Contreras}, {Coulter}, {Foley}, {Siebert}, {Morrell}, {Boutsia}, {Di Mille}, {Holoien}, {Kasen}, {Kollmeier}, {Madore}, {Monson}, {Murguia-Berthier}, {Pan}, {Prochaska}, {Ramirez-Ruiz}, {Rest}, {Adams}, {Alatalo}, {Ba{\~n}ados}, {Baughman}, {Beers}, {Bernstein}, {Bitsakis}, {Campillay}, {Hansen}, {Higgs}, {Ji}, {Maravelias}, {Marshall}, {Bidin}, {Prieto}, {Rasmussen}, {Rojas-Bravo}, {Strom}, {Ulloa}, {Vargas-Gonz{\'a}lez}, {Wan}, \& {Whitten}}]{Drout2017}
{Drout}, M.~R., {Piro}, A.~L., {Shappee}, B.~J., {et~al.} 2017, Science, 358, 1570

\bibitem[{{Fabbian} {et~al.}(2009){Fabbian}, {Asplund}, {Barklem}, {Carlsson},
  \& {Kiselman}}]{Fabbian2009}
{Fabbian}, D., {Asplund}, M., {Barklem}, P.~S., {Carlsson}, M., \& {Kiselman},
  D. 2009, \aap, 500, 1221

\bibitem[{{Feuillet} {et~al.}(2016){Feuillet}, {Bovy}, {Holtzman}, {Girardi},
  {MacDonald}, {Majewski}, \& {Nidever}}]{Feuillet2016}
{Feuillet}, D.~K., {Bovy}, J., {Holtzman}, J., {et~al.} 2016, \apj, 817, 40

\bibitem[{Foreman-Mackey(2016)}]{corner}
Foreman-Mackey, D. 2016, The Journal of Open Source Software, 24,
  doi:10.21105/joss.00024.
\newblock \url{http://dx.doi.org/10.5281/zenodo.45906}

\bibitem[{{Foreman-Mackey} {et~al.}(2013){Foreman-Mackey}, {Hogg}, {Lang}, \&
  {Goodman}}]{emcee}
{Foreman-Mackey}, D., {Hogg}, D.~W., {Lang}, D., \& {Goodman}, J. 2013, \pasp,
  125, 306

\bibitem[{{Gaia Collaboration} {et~al.}(2016{\natexlab{a}}){Gaia
  Collaboration}, {Brown}, {Vallenari}, {Prusti}, {de Bruijne}, {Mignard},
  {Drimmel}, {Babusiaux}, {Bailer-Jones}, {Bastian}, \&
  et~al.}]{GaiaCollaboration2016}
{Gaia Collaboration}, {Brown}, A.~G.~A., {Vallenari}, A., {et~al.}
  2016{\natexlab{a}}, \aap, 595, A2

\bibitem[{{Gaia Collaboration} {et~al.}(2016{\natexlab{b}}){Gaia
  Collaboration}, {Prusti}, {de Bruijne}, {Brown}, {Vallenari}, {Babusiaux},
  {Bailer-Jones}, {Bastian}, {Biermann}, {Evans}, \&
  et~al.}]{GaiaCollaboration2016a}
{Gaia Collaboration}, {Prusti}, T., {de Bruijne}, J.~H.~J., {et~al.}
  2016{\natexlab{b}}, \aap, 595, A1

\bibitem[{{Green} {et~al.}(2015){Green}, {Schlafly}, {Finkbeiner}, {Rix},
  {Martin}, {Burgett}, {Draper}, {Flewelling}, {Hodapp}, {Kaiser}, {Kudritzki},
  {Magnier}, {Metcalfe}, {Price}, {Tonry}, \& {Wainscoat}}]{Green2015}
{Green}, G.~M., {Schlafly}, E.~F., {Finkbeiner}, D.~P., {et~al.} 2015, \apj,
  810, 25

\bibitem[{{Griest} {et~al.}(2010){Griest}, {Whitmore}, {Wolfe}, {Prochaska},
  {Howk}, \& {Marcy}}]{Griest2010}
{Griest}, K., {Whitmore}, J.~B., {Wolfe}, A.~M., {et~al.} 2010, \apj, 708, 158

\bibitem[{{Hannaford} {et~al.}(1982){Hannaford}, {Lowe}, {Grevesse}, {Biemont},
  \& {Whaling}}]{Hannaford1982}
{Hannaford}, P., {Lowe}, R.~M., {Grevesse}, N., {Biemont}, E., \& {Whaling}, W.
  1982, \apj, 261, 736

\bibitem[{{Hawkins} {et~al.}(2016){Hawkins}, {Masseron}, {Jofr{\'e}},
  {Gilmore}, {Elsworth}, \& {Hekker}}]{Hawkins2016}
{Hawkins}, K., {Masseron}, T., {Jofr{\'e}}, P., {et~al.} 2016, \aap, 594, A43

\bibitem[{{Hills} \& {Day}(1976)}]{Hills1976}
{Hills}, J.~G., \& {Day}, C.~A. 1976, \aplett, 17, 87

\bibitem[{{Hinkle} {et~al.}(2000){Hinkle}, {Wallace}, {Valenti}, \&
  {Harmer}}]{Hinkle2000}
{Hinkle}, K., {Wallace}, L., {Valenti}, J., \& {Harmer}, D. 2000, {Visible and
  Near Infrared Atlas of the Arcturus Spectrum 3727-9300 A}, Vol.~2 (ASP)

\bibitem[{{Hirai} {et~al.}(2015){Hirai}, {Ishimaru}, {Saitoh}, {Fujii},
  {Hidaka}, \& {Kajino}}]{Hirai2015}
{Hirai}, Y., {Ishimaru}, Y., {Saitoh}, T.~R., {et~al.} 2015, \apj, 814, 41

\bibitem[{{Hotokezaka} {et~al.}(2018){Hotokezaka}, {Beniamini}, \&
  {Piran}}]{Hotokezaka2018}
{Hotokezaka}, K., {Beniamini}, P., \& {Piran}, T. 2018, ArXiv e-prints,
  arXiv:1801.01141

\bibitem[{Hunter(2007)}]{Matplotlib}
Hunter, J.~D. 2007, Computing in Science Engineering, 9, 90

\bibitem[{Ishimaru {et~al.}(2015)Ishimaru, Wanajo, \& Prantzos}]{Ishimaru2015}
Ishimaru, Y., Wanajo, S., \& Prantzos, S. 2015, \apjl, 804, L35

\bibitem[{{Ivans} {et~al.}(2006){Ivans}, {Simmerer}, {Sneden}, {Lawler},
  {Cowan}, {Gallino}, \& {Bisterzo}}]{Ivans2006}
{Ivans}, I.~I., {Simmerer}, J., {Sneden}, C., {et~al.} 2006, \apj, 645, 613

\bibitem[{{Izzard} {et~al.}(2018){Izzard}, {Preece}, {Jofre}, {Halabi},
  {Masseron}, \& {Tout}}]{Izzard2018}
{Izzard}, R.~G., {Preece}, H., {Jofre}, P., {et~al.} 2018, \mnras, 473, 2984

\bibitem[{{Jofr{\'e}} {et~al.}(2015){Jofr{\'e}}, {Petrucci}, {Garc{\'{\i}}a},
  \& {G{\'o}mez}}]{Jofre2015}
{Jofr{\'e}}, E., {Petrucci}, R., {Garc{\'{\i}}a}, L., \& {G{\'o}mez}, M. 2015,
  \aap, 584, L3

\bibitem[{{Jofr{\'e}} {et~al.}(2016){Jofr{\'e}}, {Jorissen}, {Van Eck},
  {Izzard}, {Masseron}, {Hawkins}, {Gilmore}, {Paladini}, {Escorza},
  {Blanco-Cuaresma}, \& {Manick}}]{Jofre2016}
{Jofr{\'e}}, P., {Jorissen}, A., {Van Eck}, S., {et~al.} 2016, \aap, 595, A60

\bibitem[{Jones {et~al.}(2001--)Jones, Oliphant, Peterson, {et~al.}}]{scipy}
Jones, E., Oliphant, T., Peterson, P., {et~al.} 2001--, {SciPy}: Open source
  scientific tools for {Python}, , , [Online; accessed <today>].
\newblock \url{http://www.scipy.org/}

\bibitem[{{Karakas} \& {Lugaro}(2016)}]{Karakas2016}
{Karakas}, A.~I., \& {Lugaro}, M. 2016, \apj, 825, 26

\bibitem[{{Kelleher} \& {Podobedova}(2008{\natexlab{a}})}]{Kelleher2008}
{Kelleher}, D.~E., \& {Podobedova}, L.~I. 2008{\natexlab{a}}, Journal of
  Physical and Chemical Reference Data, 37, 267

\bibitem[{{Kelleher} \& {Podobedova}(2008{\natexlab{b}})}]{Kelleher2008a}
---. 2008{\natexlab{b}}, Journal of Physical and Chemical Reference Data, 37,
  709

\bibitem[{{Kelleher} \& {Podobedova}(2008{\natexlab{c}})}]{Kelleher2008b}
---. 2008{\natexlab{c}}, Journal of Physical and Chemical Reference Data, 37,
  1285

\bibitem[{{Klose} {et~al.}(2002){Klose}, {Fuhr}, \& {Wiese}}]{Klose2002}
{Klose}, J.~Z., {Fuhr}, J.~R., \& {Wiese}, W.~L. 2002, Journal of Physical and
  Chemical Reference Data, 31, 217

\bibitem[{{Kock} \& {Richter}(1968)}]{Kock1968}
{Kock}, M., \& {Richter}, J. 1968, \zap, 69, 180

\bibitem[{{Lawler} {et~al.}(2001{\natexlab{a}}){Lawler}, {Bonvallet}, \&
  {Sneden}}]{Lawler2001}
{Lawler}, J.~E., {Bonvallet}, G., \& {Sneden}, C. 2001{\natexlab{a}}, \apj,
  556, 452

\bibitem[{{Lawler} \& {Dakin}(1989)}]{Lawler1989}
{Lawler}, J.~E., \& {Dakin}, J.~T. 1989, Journal of the Optical Society of
  America B Optical Physics, 6, 1457

\bibitem[{{Lawler} {et~al.}(2013){Lawler}, {Guzman}, {Wood}, {Sneden}, \&
  {Cowan}}]{Lawler2013}
{Lawler}, J.~E., {Guzman}, A., {Wood}, M.~P., {Sneden}, C., \& {Cowan}, J.~J.
  2013, \apjs, 205, 11

\bibitem[{{Lawler} {et~al.}(2015){Lawler}, {Sneden}, \& {Cowan}}]{Lawler2015}
{Lawler}, J.~E., {Sneden}, C., \& {Cowan}, J.~J. 2015, \apjs, 220, 13

\bibitem[{{Lawler} {et~al.}(2009){Lawler}, {Sneden}, {Cowan}, {Ivans}, \& {Den
  Hartog}}]{Lawler2009}
{Lawler}, J.~E., {Sneden}, C., {Cowan}, J.~J., {Ivans}, I.~I., \& {Den Hartog},
  E.~A. 2009, \apjs, 182, 51

\bibitem[{{Lawler} {et~al.}(2017){Lawler}, {Sneden}, {Nave}, {Den Hartog},
  {Emraho{\u g}lu}, \& {Cowan}}]{Lawler2017}
{Lawler}, J.~E., {Sneden}, C., {Nave}, G., {et~al.} 2017, \apjs, 228, 10

\bibitem[{{Lawler} {et~al.}(2001{\natexlab{b}}){Lawler}, {Wickliffe}, {den
  Hartog}, \& {Sneden}}]{Lawler2001a}
{Lawler}, J.~E., {Wickliffe}, M.~E., {den Hartog}, E.~A., \& {Sneden}, C.
  2001{\natexlab{b}}, \apj, 563, 1075

\bibitem[{{Lawler} {et~al.}(2014){Lawler}, {Wood}, {Den Hartog}, {Feigenson},
  {Sneden}, \& {Cowan}}]{Lawler2014}
{Lawler}, J.~E., {Wood}, M.~P., {Den Hartog}, E.~A., {et~al.} 2014, \apjs, 215,
  20

\bibitem[{{Luck}(1994)}]{Luck1994}
{Luck}, R.~E. 1994, \apjs, 91, 309

\bibitem[{{Majewski} {et~al.}(2017){Majewski}, {Schiavon}, {Frinchaboy},
  {Allende Prieto}, {Barkhouser}, {Bizyaev}, {Blank}, {Brunner}, {Burton},
  {Carrera}, {Chojnowski}, {Cunha}, {Epstein}, {Fitzgerald}, {Garc{\'{\i}}a
  P{\'e}rez}, {Hearty}, {Henderson}, {Holtzman}, {Johnson}, {Lam}, {Lawler},
  {Maseman}, {M{\'e}sz{\'a}ros}, {Nelson}, {Nguyen}, {Nidever}, {Pinsonneault},
  {Shetrone}, {Smee}, {Smith}, {Stolberg}, {Skrutskie}, {Walker}, {Wilson},
  {Zasowski}, {Anders}, {Basu}, {Beland}, {Blanton}, {Bovy}, {Brownstein},
  {Carlberg}, {Chaplin}, {Chiappini}, {Eisenstein}, {Elsworth}, {Feuillet},
  {Fleming}, {Galbraith-Frew}, {Garc{\'{\i}}a}, {Garc{\'{\i}}a-Hern{\'a}ndez},
  {Gillespie}, {Girardi}, {Gunn}, {Hasselquist}, {Hayden}, {Hekker}, {Ivans},
  {Kinemuchi}, {Klaene}, {Mahadevan}, {Mathur}, {Mosser}, {Muna}, {Munn},
  {Nichol}, {O'Connell}, {Parejko}, {Robin}, {Rocha-Pinto}, {Schultheis},
  {Serenelli}, {Shane}, {Silva Aguirre}, {Sobeck}, {Thompson}, {Troup},
  {Weinberg}, \& {Zamora}}]{Majewski2017}
{Majewski}, S.~R., {Schiavon}, R.~P., {Frinchaboy}, P.~M., {et~al.} 2017, \aj,
  154, 94

\bibitem[{{Martig} {et~al.}(2015){Martig}, {Rix}, {Aguirre}, {Hekker},
  {Mosser}, {Elsworth}, {Bovy}, {Stello}, {Anders}, {Garc{\'{\i}}a}, {Tayar},
  {Rodrigues}, {Basu}, {Carrera}, {Ceillier}, {Chaplin}, {Chiappini},
  {Frinchaboy}, {Garc{\'{\i}}a-Hern{\'a}ndez}, {Hearty}, {Holtzman}, {Johnson},
  {Majewski}, {Mathur}, {M{\'e}sz{\'a}ros}, {Miglio}, {Nidever}, {Pan},
  {Pinsonneault}, {Schiavon}, {Schneider}, {Serenelli}, {Shetrone}, \&
  {Zamora}}]{Martig2015a}
{Martig}, M., {Rix}, H.-W., {Aguirre}, V.~S., {et~al.} 2015, \mnras, 451, 2230

\bibitem[{{Matteucci} \& {Greggio}(1986)}]{Matteucci1986}
{Matteucci}, F., \& {Greggio}, L. 1986, \aap, 154, 279

\bibitem[{{McCrea}(1964)}]{McCrea1964}
{McCrea}, W.~H. 1964, \mnras, 128, 147

\bibitem[{Mckinney(2010)}]{pandas}
Mckinney, W. 2010

\bibitem[{{McWilliam}(1998)}]{McWilliam1998}
{McWilliam}, A. 1998, \aj, 115, 1640

\bibitem[{{Mel{\'e}ndez} \& {Barbuy}(2009)}]{Melendez2009}
{Mel{\'e}ndez}, J., \& {Barbuy}, B. 2009, \aap, 497, 611

\bibitem[{{Miglio} {et~al.}(2013){Miglio}, {Chiappini}, {Morel}, {Barbieri},
  {Chaplin}, {Girardi}, {Montalb{\'a}n}, {Valentini}, {Mosser}, {Baudin},
  {Casagrande}, {Fossati}, {Silva Aguirre}, \& {Baglin}}]{Miglio2013}
{Miglio}, A., {Chiappini}, C., {Morel}, T., {et~al.} 2013, \mnras, 429, 423

\bibitem[{{Nandy} {et~al.}(2012){Nandy}, {Singh}, {Shah}, \&
  {Sahoo}}]{Nandy2012}
{Nandy}, D.~K., {Singh}, Y., {Shah}, B.~P., \& {Sahoo}, B.~K. 2012, \pra, 86,
  052517

\bibitem[{{Nishimura} {et~al.}(2017){Nishimura}, {Sawai}, {Takiwaki}, {Yamada}, 
     \& {Thielemann}}]{Nishimura2017}
{Nishimura}, N., {Sawai}, H., {Takiwaki}, T., {Yamada}, S., \& 
	{Thielemann}, F.-K. 2017, \apjl, 836, L21 

\bibitem[{{O'Brian} {et~al.}(1991){O'Brian}, {Wickliffe}, {Lawler}, {Whaling},
  \& {Brault}}]{OBrian1991}
{O'Brian}, T.~R., {Wickliffe}, M.~E., {Lawler}, J.~E., {Whaling}, W., \&
  {Brault}, J.~W. 1991, Journal of the Optical Society of America B Optical
  Physics, 8, 1185

\bibitem[{{Pehlivan Rhodin} {et~al.}(2017){Pehlivan Rhodin}, {Hartman},
  {Nilsson}, \& {J{\"o}nsson}}]{Rhodin2017}
{Pehlivan Rhodin}, A., {Hartman}, H., {Nilsson}, H., \& {J{\"o}nsson}, P. 2017,
  \aap, 598, A102

\bibitem[{Press {et~al.}(1992)Press, Flannery, Teukolsky, \&
  Vetterling}]{Press1992}
Press, W.~H., Flannery, B.~P., Teukolsky, S.~A., \& Vetterling, W.~T. 1992,
  {Numerical Recipes in Fortran 77: The Art of Scientific Computing}, 2nd edn.
  (Cambridge University Press)

\bibitem[{{Ram{\'{\i}}rez} {et~al.}(2007){Ram{\'{\i}}rez}, {Allende Prieto}, \&
  {Lambert}}]{Ramirez2007}
{Ram{\'{\i}}rez}, I., {Allende Prieto}, C., \& {Lambert}, D.~L. 2007, \aap,
  465, 271

\bibitem[{{Ram{\'{\i}}rez} {et~al.}(2014){Ram{\'{\i}}rez}, {Mel{\'e}ndez},
  {Bean}, {Asplund}, {Bedell}, {Monroe}, {Casagrande}, {Schirbel}, {Dreizler},
  {Teske}, {Tucci Maia}, {Alves-Brito}, \& {Baumann}}]{Ramirez2014}
{Ram{\'{\i}}rez}, I., {Mel{\'e}ndez}, J., {Bean}, J., {et~al.} 2014, \aap, 572,
  A48

\bibitem[{{Rosman} \& {Taylor}(1998)}]{Rosman1998}
{Rosman}, K.~J.~R., \& {Taylor}, P.~D.~P. 1998, Journal of Physical and
  Chemical Reference Data, 27, 1275

\bibitem[{{Ruffoni} {et~al.}(2014){Ruffoni}, {Den Hartog}, {Lawler}, {Brewer},
  {Lind}, {Nave}, \& {Pickering}}]{Ruffoni2014}
{Ruffoni}, M.~P., {Den Hartog}, E.~A., {Lawler}, J.~E., {et~al.} 2014, \mnras,
  441, 3127

\bibitem[{{Smith}(1988)}]{Smith1988}
{Smith}, G. 1988, Journal of Physics B Atomic Molecular Physics, 21, 2827

\bibitem[{{Smith} \& {Raggett}(1981)}]{Smith1981}
{Smith}, G., \& {Raggett}, D.~S.~J. 1981, Journal of Physics B Atomic Molecular
  Physics, 14, 4015

\bibitem[{{Sneden}(1973)}]{Sneden1973}
{Sneden}, C. 1973, \apj, 184, 839

\bibitem[{{Sneden} {et~al.}(2008){Sneden}, {Cowan}, \& {Gallino}}]{Sneden2008}
{Sneden}, C., {Cowan}, J.~J., \& {Gallino}, R. 2008, \araa, 46, 241

\bibitem[{{Sobeck} {et~al.}(2007){Sobeck}, {Lawler}, \& {Sneden}}]{Sobeck2007}
{Sobeck}, J.~S., {Lawler}, J.~E., \& {Sneden}, C. 2007, \apj, 667, 1267

\bibitem[{{Soderblom}(2010)}]{Soderblom2010}
{Soderblom}, D.~R. 2010, \araa, 48, 581

\bibitem[{{Spina} {et~al.}(2018){Spina}, {Mel{\'e}ndez}, {Karakas}, {dos
  Santos}, {Bedell}, {Asplund}, {Ram{\'{\i}}rez}, {Yong}, {Alves-Brito},
  {Bean}, \& {Dreizler}}]{Spina2018}
{Spina}, L., {Mel{\'e}ndez}, J., {Karakas}, A.~I., {et~al.} 2018, \mnras, 474,
  2580

\bibitem[{{Takeda} \& {Tajitsu}(2017)}]{Takeda2017}
{Takeda}, Y., \& {Tajitsu}, A. 2017, \pasj, 69, 74

\bibitem[{{Tinsley}(1979)}]{Tinsley1979}
{Tinsley}, B.~M. 1979, \apj, 229, 1046

\bibitem[{van~der Walt {et~al.}(2011)van~der Walt, Colbert, \&
  Varoquaux}]{numpy}
van~der Walt, S., Colbert, S.~C., \& Varoquaux, G. 2011, Computing in Science
  \& Engineering, 13, 22.
\newblock \url{http://aip.scitation.org/doi/abs/10.1109/MCSE.2011.37}

\bibitem[{{Vogt} {et~al.}(1994){Vogt}, {Allen}, {Bigelow}, {Bresee}, {Brown},
  {Cantrall}, {Conrad}, {Couture}, {Delaney}, {Epps}, {Hilyard}, {Hilyard},
  {Horn}, {Jern}, {Kanto}, {Keane}, {Kibrick}, {Lewis}, {Osborne},
  {Pardeilhan}, {Pfister}, {Ricketts}, {Robinson}, {Stover}, {Tucker}, {Ward},
  \& {Wei}}]{Vogt1994}
{Vogt}, S.~S., {Allen}, S.~L., {Bigelow}, B.~C., {et~al.} 1994, in \procspie,
  Vol. 2198, Instrumentation in Astronomy VIII, ed. D.~L. {Crawford} \& E.~R.
  {Craine}, 362

\bibitem[{{Wanajo} {et~al.}(2014){Wanajo}, {Sekiguchi}, {Nishimura}, {Kiuchi},
  {Kyutoku}, \& {Shibata}}]{Wanajo2014}
{Wanajo}, S., {Sekiguchi}, Y., {Nishimura}, N., {et~al.} 2014, \apjl, 789, L39

\bibitem[{{Wood} {et~al.}(2013){Wood}, {Lawler}, {Sneden}, \&
  {Cowan}}]{Wood2013}
{Wood}, M.~P., {Lawler}, J.~E., {Sneden}, C., \& {Cowan}, J.~J. 2013, \apjs,
  208, 27

\bibitem[{{Wood} {et~al.}(2014){Wood}, {Lawler}, {Sneden}, \&
  {Cowan}}]{Wood2014}
---. 2014, \apjs, 211, 20

\bibitem[{{Yong} {et~al.}(2016){Yong}, {Casagrande}, {Venn}, {Chen{\'e}},
  {Keown}, {Malo}, {Martioli}, {Alves-Brito}, {Asplund}, {Dotter}, {Martell},
  {Mel{\'e}ndez}, \& {Schlesinger}}]{Yong2016}
{Yong}, D., {Casagrande}, L., {Venn}, K.~A., {et~al.} 2016, \mnras, 459, 487

\bibitem[{{York} {et~al.}(2000){York}, {Adelman}, {Anderson}, {Anderson},
  {Annis}, {Bahcall}, {Bakken}, {Barkhouser}, {Bastian}, {Berman}, {Boroski},
  {Bracker}, {Briegel}, {Briggs}, {Brinkmann}, {Brunner}, {Burles}, {Carey},
  {Carr}, {Castander}, {Chen}, {Colestock}, {Connolly}, {Crocker}, {Csabai},
  {Czarapata}, {Davis}, {Doi}, {Dombeck}, {Eisenstein}, {Ellman}, {Elms},
  {Evans}, {Fan}, {Federwitz}, {Fiscelli}, {Friedman}, {Frieman}, {Fukugita},
  {Gillespie}, {Gunn}, {Gurbani}, {de Haas}, {Haldeman}, {Harris}, {Hayes},
  {Heckman}, {Hennessy}, {Hindsley}, {Holm}, {Holmgren}, {Huang}, {Hull},
  {Husby}, {Ichikawa}, {Ichikawa}, {Ivezi{\'c}}, {Kent}, {Kim}, {Kinney},
  {Klaene}, {Kleinman}, {Kleinman}, {Knapp}, {Korienek}, {Kron}, {Kunszt},
  {Lamb}, {Lee}, {Leger}, {Limmongkol}, {Lindenmeyer}, {Long}, {Loomis},
  {Loveday}, {Lucinio}, {Lupton}, {MacKinnon}, {Mannery}, {Mantsch}, {Margon},
  {McGehee}, {McKay}, {Meiksin}, {Merelli}, {Monet}, {Munn}, {Narayanan},
  {Nash}, {Neilsen}, {Neswold}, {Newberg}, {Nichol}, {Nicinski}, {Nonino},
  {Okada}, {Okamura}, {Ostriker}, {Owen}, {Pauls}, {Peoples}, {Peterson},
  {Petravick}, {Pier}, {Pope}, {Pordes}, {Prosapio}, {Rechenmacher}, {Quinn},
  {Richards}, {Richmond}, {Rivetta}, {Rockosi}, {Ruthmansdorfer}, {Sandford},
  {Schlegel}, {Schneider}, {Sekiguchi}, {Sergey}, {Shimasaku}, {Siegmund},
  {Smee}, {Smith}, {Snedden}, {Stone}, {Stoughton}, {Strauss}, {Stubbs},
  {SubbaRao}, {Szalay}, {Szapudi}, {Szokoly}, {Thakar}, {Tremonti}, {Tucker},
  {Uomoto}, {Vanden Berk}, {Vogeley}, {Waddell}, {Wang}, {Watanabe},
  {Weinberg}, {Yanny}, {Yasuda}, \& {SDSS Collaboration}}]{York2000}
{York}, D.~G., {Adelman}, J., {Anderson}, Jr., J.~E., {et~al.} 2000, \aj, 120,
  1579

\end{thebibliography}

\end{document}